\documentclass[11pt,final,conference,onecolumn]{IEEEtran}
\IEEEoverridecommandlockouts
\usepackage{graphicx}
\usepackage{multirow}
\usepackage{subcaption}
\usepackage{caption}
\usepackage{url}
\usepackage{amsmath,amssymb,amsfonts}
\usepackage{algorithmic}
\usepackage{algorithm}
\usepackage{textcomp}
\usepackage{xcolor}
\usepackage{enumitem}
\definecolor{darkgreen}{RGB}{0,194,0}

\def\vector#1{\mbox{\boldmath $#1$}}
\def\BibTeX{{\rm B\kern-.05em{\sc i\kern-.025em b}\kern-.08em
    T\kern-.1667em\lower.7ex\hbox{E}\kern-.125emX}}
\begin{document}

\title{Speech privacy-preserving methods using secret key for convolutional neural network models and their robustness evaluation}

\author{\IEEEauthorblockN{1\textsuperscript{st} Shoko Niwa}
\IEEEauthorblockA{Tokyo Metropolitan University\\
Tokyo,Japan \\
niwa-shoko@ed.tmu.ac.jp}
\and
\IEEEauthorblockN{2\textsuperscript{nd} Sayaka Shiota}
\IEEEauthorblockA{Tokyo Metropolitan University\\
Tokyo,Japan  \\
sayaka@tmu.ac.jp}
\and 
\IEEEauthorblockN{3\textsuperscript{rd} Hitoshi Kiya}
\IEEEauthorblockA{Tokyo Metropolitan University\\
Tokyo,Japan \\
kiya@tmu.ac.jp}

}

\maketitle

\begin{abstract}
In this paper, we propose privacy-preserving methods with a secret key for convolutional neural network (CNN)-based models in speech processing tasks. 
In environments where untrusted third parties, like cloud servers, provide CNN-based systems, ensuring the privacy of speech queries becomes essential. 
This paper proposes encryption methods for speech queries using secret keys and a model structure that allows for encrypted queries to be accepted without decryption.
Our approach introduces three types of secret keys: Shuffling, Flipping, and random orthogonal matrix (ROM).
In experiments, we demonstrate that when the proposed methods are used with the correct key, identification performance did not degrade. 
Conversely, when an incorrect key is used, the performance significantly decreased. 
Particularly, with the use of ROM, we show that even with a relatively small key space, high privacy-preserving performance can be maintained many speech processing tasks. Furthermore, we also demonstrate the difficulty of recovering original speech from encrypted queries in various robustness evaluations.
\end{abstract}


\section{Introduction}\label{sec:intro}
Speech data usually includes personal information such as age, gender, language, and speaking content.
To exploit such information, CNN models for tasks such as automatic speech recognition, speech synthesis, and speaker verification have been actively studied~\cite{Watanabe2018, yang21c_interspeech}.
In recent years, CNN models and speech data are increasingly uploaded to or stored on cloud servers via the Internet, and CNN models are run on cloud servers.
However, since cloud services are managed by external providers, various threats such as data leakage due to malicious attacks from outside or inside are a concern~\cite{tabrizchi2020survey,SINGH201788}.
When using CNN models on a cloud service, it is necessary to provide a trained model and query data to the cloud service. 
Therefore, when cloud services are insecure, models and queries face threats. 
To prevent such risks, it is important to preserve privacy before sending data to insecure services.
The issue of privacy has been gradually gaining attention as the latest topic in the research field of speech processing~\cite{tomashenko2022voiceprivacy,KAI2022101315,kai22_odyssey, 8683721, teixeira22_interspeech}.
However, most of the existing methods for preserving the privacy of speech focus on concealing information about the speaker of the speech~\cite{tomashenko2022voiceprivacy}, and little is mentioned about concealing the content of speech~\cite{williams22_spsc}.
There are also problems such as model performance degradation when the existing methods are in use. 

In the research field of image processing, many privacy-preserving methods have been proposed for CNN-based systems~\cite{kiya2022overview, maungprivacy}.
These methods propose a framework wherein models can process encrypted images using a secret key without the need for decryption, thereby protecting the visual information.
Inspired by the research, we have proposed a simple privacy-preserving method using a secret key for speech data~\cite{niwa2023_eusipco}. 
This approach regarded the privacy-preserving of speech data as protection of the auditory information. Thus, this research on the privacy-preserving of speech data aimed to control the performance of speech processing systems by using a secret key. This paper expands the robustness of this initial research and presents the robustness evaluation for speech privacy-preserving. 
\textcolor{black}{The privacy-preserving methods used in this paper, e.g. Shuffling, Flipping, ROM are common in biometric template protection~\cite{KONG20061359, LUMINI20071057} and privacy-preserving image classification areas. 
Our contributions are not only to apply them to speech areas but to also propose the method that can avoid the performance degradation of models.}
The proposed method encrypts the speech queries with secret keys and uses a model structure that allows encrypted speech queries to be accepted without decryption. To realize this framework,
we assume that the first layer of the CNN model is a convolutional layer, and that the kernel size and stride size of the first convolutional layer are equal.
For model encryption, the kernel of the first convolutional layer of the model is encrypted using a matrix corresponding to the matrix used as the secret key. 
This operation allows encrypted speech data to be input directly into the model without decryption, as it cancels out the effect of encryption on the input speech data.
Our approach introduces three types of secret keys: permutation matrix, sign inversion, and random orthogonal matrix.
To validate the advantages of our approach, which include task independence and the absence of the need to retrain the model as long as certain conditions are met, we conducted performance evaluations of our privacy-preserving methods using three tasks: automatic speaker verification~(ASV), automatic speech recognition~(ASR), and acoustic scene classification~(ASC) task.
The experimental results show that CNN models can be used with the same accuracy as before encryption, when speech encrypted with the correct secret key is input to the model encrypted with the correct secret key.
It is also shown that the accuracies of the CNN models are significantly reduced when the input speech is encrypted with an incorrect secret key.
In particular, it is shown that using a random orthogonal matrix as a secret key can preserve speech privacy while maintaining a large key space, even when the block size is small.
Furthermore, experiments done to evaluate robustness show that when audio encrypted with an incorrect secret key is input to a model encrypted with the correct secret key, the performance of the model decreases steadily for larger block sizes, and stable privacy-preserving performance is obtained.
In addition, it is shown that speech data encrypted with the proposed methods cannot be reconstructed unless the correct secret key is used\footnote{\textcolor{black}{The code used to generate the secret key, encrypt the model, and encrypt query speech data is available at \url{https://github.com/kiyalab-tmu/SecretKeyVoicePrivacyPreserving-CNN}}}.

The following is the structure of the paper.
In Section~\ref{sec:image}, we describe the privacy-preserving scenario that we assume. 
Section~\ref{sec:proposed} gives the details on the proposed methods and in Section~\ref{seq:experiment}, we show the experimental results. 
In Section~\ref{seq:conclude}, we conclude the study and describe our future work.

\begin{figure}[t]
    \centering
    \includegraphics[keepaspectratio,width=11cm]{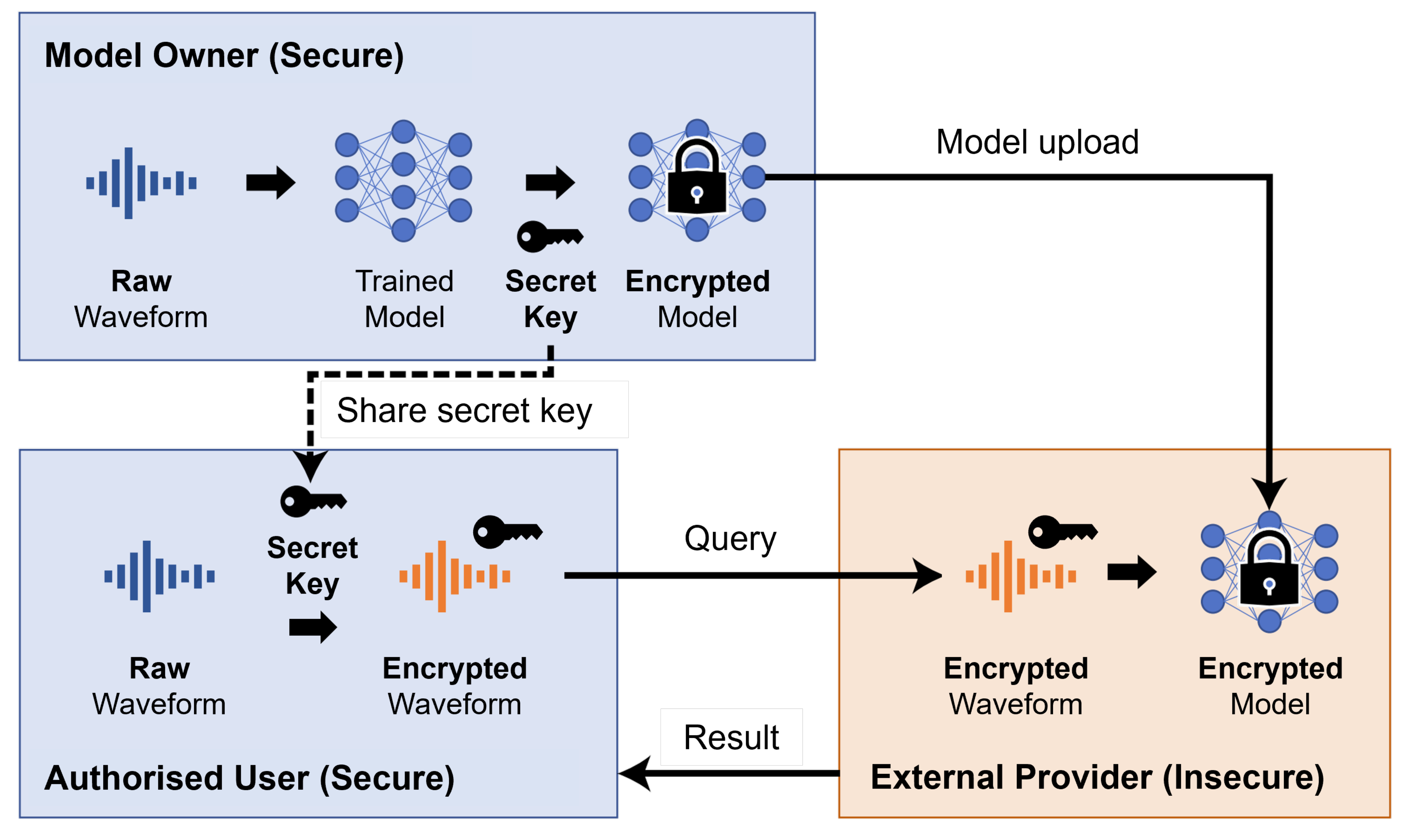}
    \caption{Privacy-preserving scenario}
    \label{fig:gaiyou}
\end{figure}

\section{Privacy-preserving scenario}\label{sec:image}
The scenario assumed for the privacy-preserving frameworks in this paper is illustrated in Fig.~\ref{fig:gaiyou}.
Figure~\ref{fig:gaiyou} consists of the model owner, the authorized user, and the external provider. The external provider is assumed to be untrusted, while the model owner and the authorized user are considered secure.
First, the model owner trains a CNN model to process plain speech data, such as spectrograms and waveforms, within a secure environment. 
The trained model is encrypted with a secret key. 
Subsequently, the model owner provides the encrypted model to an external provider, such as a cloud service, and shares the secret key used for the model encryption with an authorized user.
It is assumed that the external provider, being managed by a third party, is not within a secure environment.
When the authorized user utilizes the encrypted CNN model from the external provider, the user encrypts the query speech data using the secret key received from the model owner and uploads the encrypted speech data to the external provider.
Finally, the external provider inputs the encrypted speech data into the encrypted model and returns the result to the authorized user.
In this scenario, even if the external provider is not secure, only authorized users possessing the correct secret key can utilize the encrypted model as intended by the model owner,
and only encrypted speech, with the privacy information concealed, is stored with the external provider. This ensures that privacy is maintained even in the event of a data leakage.

Similar to this task, the VoicePrivacy challenge~\cite{tomashenko2022voiceprivacy} is famously related to speech privacy in the speech processing area. However, the privacy-preserving scenario of the VoicePrivacy challenge considers that the speaker wishes to keep their identity confidential while not protecting the content of the speech. In this case, the attacker aims to identify the speaker from the speech data. On the other hand, our paper presents a scenario where the user encrypts the raw speech data to preserve personal information, including the speech's content or a speaker's information. The attacker aims to steal and misuse speech data. Our method conceals auditory information by changing the raw speech data with a secret key. 

\section{Proposed methods}\label{sec:proposed}
This section describes the encryption methods for query speech data, the methods for model encryption, and the advantage of using an orthogonal matrix as the secret key in the proposed methods.
The proposed methods are familiar in the image processing area. However, we show the theoretical correctness of adapting the framework to speech data, which are represented in one or two dimensions.

\subsection{Query encryption}\label{sec:queryenc}
We present a procedure for encrypting speech data using a secret key.
Speech data consists of variable-length one-dimensional data, such as waveforms, or variable-length two-dimensional data, such as spectrograms obtained by applying a short-time Fourier transform to waveforms.
Here, we propose three encryption methods using three kinds of orthogonal matrix, that is, the Shuffling in \ref{sec:pix-speech-enc}, Flipping in \ref{sec:bit-speech-enc}, and random orthogonal matrix (ROM) in \ref{sec:orth-speech-enc}, which are detailed below.

\begin{figure}[tp]
\begin{algorithm}[H]
    \caption{Shuffling speech data encryption}
    \label{alg:ps}
    \begin{algorithmic}[1] 
    \REQUIRE $\vector{X}\text{: speech data}$, $M\text{: block size}$, $\vector{K}_\text{s}\text{: secret key}$
    \ENSURE $\vector{X}^{(\boldsymbol{K}_\text{s})}$: encrypted speech data
    \STATE $\vector{X}^{(\boldsymbol{K}_\text{s})} \leftarrow O$
    \STATE $f \leftarrow [F/M]$
    \STATE $t \leftarrow [T/M]$
    \FOR{$((i=0; i<f; i=i+M))$}
    \FOR{$((j=0; j<t; j=j+M))$}
    \STATE $\vector{X}_{ij} \leftarrow \vector{X}[i:i+M, j:j+M]$
    \STATE $\hat{\vector{X}}_{ij} \leftarrow \mathrm{flatten}(\vector{X}_{ij})$
    \STATE $\hat{\vector{X}}_{ij}^{(\boldsymbol{K}_\text{s})} \leftarrow \hat{\vector{X}}_{ij}[\vector{K}_\text{s}]$
    \STATE $\vector{X}_{ij}^{(\boldsymbol{K}_\text{s})} \leftarrow \mathrm{reshape}(\hat{\vector{X}}_{ij}^{(\boldsymbol{K}_\text{s})})$
    \STATE $\vector{X}^{(\boldsymbol{K}_\text{s})}[i:i+M, j:j+M] \leftarrow \vector{X}_{ij}^{(\boldsymbol{K}_\text{s})})$
    \ENDFOR
    \ENDFOR
    \RETURN $\vector{X}^{(\boldsymbol{K}_\text{s})}$
    \end{algorithmic}
\end{algorithm}
\end{figure}

\begin{figure}[tp]
\begin{algorithm}[H]
    \caption{Secret key generation in Shuffling}
    \label{alg:scr-ps}
    \begin{algorithmic}[1] 
    \REQUIRE $M$: block size 
\ENSURE $\vector{K}_\text{s}$: secret key
\STATE $\vector{K}_\text{s} \leftarrow [0, 1, \ldots, M-1]$
\FOR {$i = M-1$ to $1$}
    \STATE $j \gets$ random integer from $0$ to $i$
    \STATE Swap $\vector{K}_\text{s}[i]$ and $\vector{K}_\text{s}[j]$
\ENDFOR
\STATE \RETURN $\vector{K}_\text{s}$
    \end{algorithmic}
\end{algorithm}
\end{figure}

\subsubsection{Shuffling}\label{sec:pix-speech-enc}
In this subsection, we describe an encryption method for speech data using ``Shuffling'', which uses a permutation matrix as the secret key.
Algorithm~\ref{alg:ps} outlines the algorithm for encrypting speech data using Shuffling.
\begin{enumerate}
\item Speech data~$\vector{X}$ is divided into blocks of block size~$M$ as follows:
\begin{equation}\label{eq:whole}
\vector{X}=
\begin{bmatrix}
\vector{X}_{11}   &  \dots  &  \vector{X}_{1j}  &  \dots  &  \vector{X}_{1t}\\
\vdots   &      &  \vdots    &    &     \vdots  \\
\vector{X}_{i1}    &   \dots  &  \vector{X}_{ij}  &  \dots  &  \vector{X}_{it}\\
\vdots   &      &  \vdots    &    &     \vdots  \\
\vector{X}_{f1}    &   \dots  &  \vector{X}_{fj}  &  \dots  &  \vector{X}_{ft}\\
\end{bmatrix}.
\end{equation}
When defining the size of the speech data~$\vector{X}$ as an $F\times T$ matrix, $F$ represents the size in the frequency direction, and $T$ represents the size in the time direction. 
For one-dimensional data such as a speech waveform, $F$ is set to $1$, while $T$ is set to $[T/M]$, where $M$ is the block size.
For two-dimensional data like a spectrogram, $F$ is determined by the frequency resolution, while both $T$ and $F$ are set to $[T/M]$ and $[F/M]$, respectively.
Each block~$\vector{X}_{ij}$ within the matrix $\vector{X}$ is further defined as follows:
\begin{equation}\label{eq:block}
\vector{X}_{ij}=
\begin{bmatrix}
x_{11}    &  x_{12}   &  \dots   &  x_{1m}\\
x_{21}    &  x_{22}   &  \dots   &  x_{2m}\\
\vdots   &   \vdots   &    &       \vdots  \\
x_{n1}    &  x_{n2}   &  \dots   &  x_{nm}\\
\end{bmatrix}.
\end{equation}
The size of the block~$\vector{X}_{ij}$ can be expressed as $n\times m$, when $\vector{X}$ represents one-dimensional data, $n=1$ and $m=M$, and when $\vector{X}$ represents two-dimensional data, $n$ and $m$ are both set to $M$.

\item To encrypt each block~$\vector{X}_{ij}$ using a secret key, $\vector{X}_{ij}$ is flattened to a one-dimensional vector~$\hat{\vector{X}}_{ij}$ as follows:
\begin{eqnarray}
\hat{\vector{X}}_{ij}   & =  &  
\mathrm{flatten}(\vector{X}_{ij}) \nonumber \\
  & =  & 
\left[ 
{x}_{1}, {x}_{2}, \dots, {x}_{m}, {x}_{m+1}, \dots, {x}_{N}
\right],
\label{eq:flatten}
\end{eqnarray}
where $N$ denotes the total number of elements in the block $\vector{X}_{ij}$, i.e., $N=n\times m$.
The $\mathrm{flatten}$ function re-indexes the index of each element~$x$ as a row vector and behaves as follows: $[{x}_{1}, {x}_{2}, \dots, {x}_{m}, $
${x}_{m+1}, \dots, {x}_{N}]=[x_{11}, x_{12}, \dots, x_{1m}, x_{21}, \dots, x_{nm}]$.

\item The secret key~$\vector{K}_\text{s}$ is generated as follows:
\begin{equation}\label{eq:s}
\vector{K}_\text{s}=\{ k_\text{s}(1), k_\text{s}(2), \dots, k_\text{s}(k), \dots, k_\text{s}(N)\},
\end{equation}
where the symbols $l$ and $k$ in Eq.~\eqref{eq:s} are indices of $k_s$, $k_\text{s}(k)\in \{ 1,2,\dots, N\}$, $k_\text{s}(k)\ne k_\text{s}(l)$, $k,l \in \{ 1,2, \dots, N\}$, $k\ne l$.

\item The permutation matrix~$\vector{K}'_\text{s}$ used for encryption is generated as follows:
\begin{equation}\label{eq:key_s}
\vector{K}'_\text{s}=[ \vector{e}_{k_\text{s}(1)}, \vector{e}_{k_\text{s}(2)}, \dots, \vector{e}_{k_\text{s}(k)}, \dots, \vector{e}_{k_\text{s}(N)} ].
\end{equation}
Let $\vector{e}_{k_\text{s}(i)}$ denote the unit vector.

\item The matrix product of the one-dimensional vector~$\hat{\vector{X}}_{ij}$ and the permutation matrix~$\vector{K}'_\text{s}$ is calculated to obtain the encrypted row vector~$\hat{\vector{X}}_{ij}^{(\boldsymbol{K}_\text{s})}$ as follows: 
\begin{eqnarray}
\hat{\vector{X}}_{ij}^{(\boldsymbol{K}_\text{s})t}   & =  &  
\vector{K}'_\text{s}\hat{\vector{X}}_{ij}^t \nonumber\\
  & =  &  
\left[ \vector{e}_{s(1)}, \dots, \vector{e}_{s(N)}  \right] \left[ {x}_1, \dots, {x}_N \right]^t\nonumber\\
  & =  & 
\left[ 
{x}^{(k_\text{s})}_{1}, \dots, {x}^{(k_\text{s})}_{N}
\right] ^t.
\label{eq:enc_q_s}
\end{eqnarray}

\item Using the $\mathrm{reshape}$ function, the one-dimensional vector~$\hat{\vector{X}}_{ij}^{(\boldsymbol{K}_\text{s})}$ is transformed to a matrix of the same shape as the block~$\vector{X}_{ij}$ to obtain an encrypted block~$\vector{X}_{ij}^{(\boldsymbol{K}_\text{s})}$ as follows:
\begin{eqnarray}
\vector{X}_{ij}^{(\boldsymbol{K}_\text{s})}   & =  &  
\mathrm{reshape}(\hat{\vector{X}}_{ij}^{(\boldsymbol{K}_\text{s})}) \nonumber \\
  & =  & 
\begin{bmatrix}
x^{(k_\text{s})}_{11}    &  x^{(k_\text{s})}_{12}   &  \dots   &  x^{(k_\text{s})}_{1m}\\
x^{(k_\text{s})}_{21}    &  x^{(k_\text{s})}_{22}   &  \dots   &  x^{(k_\text{s})}_{2m}\\
\vdots   &   \vdots   &    &       \vdots  \\
x^{(k_\text{s})}_{n1}    &  x^{(k_\text{s})}_{n2}   &  \dots   &  x^{(k_\text{s})}_{nm}\\
\end{bmatrix}.
\label{eq:reshape}
\end{eqnarray}

\item All the blocks of the speech data~$\vector{X}$ that were divided in step $1$ are processed in steps $2$ - $6$ to obtain the encrypted speech data~$\vector{X}^{(\boldsymbol{K}_\text{s})}$.
$\vector{X}^{(\boldsymbol{K}_\text{s})}$ can be expressed as follows:
\begin{equation}\label{eq:encx}
\vector{X}^{(\boldsymbol{K}_\text{s})}=
\begin{bmatrix}
\vector{X}^{(\boldsymbol{K}_\text{s})}_{11}    &  \dots   &  \vector{X}^{(\boldsymbol{K}_\text{s})}_{1t}\\
\vdots   &    \ddots   &       \vdots  \\
\vector{X}^{(\boldsymbol{K}_\text{s})}_{f1}    &  \dots   &  \vector{X}^{(\boldsymbol{K}_\text{s})}_{ft}\\
\end{bmatrix}.
\end{equation}
\end{enumerate}

The steps from 1 to 7 demonstrate the transformation of plain speech data into encrypted speech data with the secret key $\vector{K}_\text{s}$.

\begin{figure}[tp]
\begin{algorithm}[H]
    \caption{Flipping speech data encryption}
    \label{alg:bf}
    \begin{algorithmic}[1] 
    \REQUIRE $\vector{X}\text{: speech data}$, $M\text{: block size}$, $\vector{K}_\text{f}\text{: secret key}$
    \ENSURE $\vector{X}^{(\boldsymbol{K}_\text{f})}$: encrypted speech data
    \STATE $\vector{X}^{(\boldsymbol{K}_\text{f})} \leftarrow O$
    \STATE $f \leftarrow [F/M]$
    \STATE $t \leftarrow [T/M]$
    \FOR{$((i=0; i<f; i=i+M))$}
    \FOR{$((j=0; j<t; j=j+M))$}
    \STATE $\vector{X}_{ij} \leftarrow \vector{X}[i:i+M, j:j+M]$
    \STATE $\hat{\vector{X}}_{ij} \leftarrow \mathrm{flatten}(\vector{X}_{ij})$
    \IF{$\vector{K}_\text{f}[j]=1$}
    \STATE $\hat{\vector{X}}_{ij}^{(\boldsymbol{K}_\text{f})}[j] \leftarrow -\hat{\vector{X}}_{ij}[j]$
    \ENDIF
    \STATE $\vector{X}_{ij}^{(\boldsymbol{K}_\text{f})} \leftarrow \mathrm{reshape}(\hat{\vector{X}}_{ij}^{(\boldsymbol{K}_\text{f})})$
    \STATE $\vector{X}^{(\boldsymbol{K}_\text{f})}[i:i+M, j:j+M] \leftarrow \vector{X}_{ij}^{(\boldsymbol{K}_\text{f})}$
    \ENDFOR
    \ENDFOR
    \RETURN $\vector{X}^{(\boldsymbol{K}_\text{f})}$
    \end{algorithmic}
\end{algorithm}
\end{figure}

\begin{figure}[tp]
\begin{algorithm}[H]
    \caption{Secret key generation in Flipping}
    \label{alg:scr-bf}
    \begin{algorithmic}[1] 
    \REQUIRE $M$: block size 
\ENSURE $\vector{K}_\text{f}$: secret key
\STATE Initialize $\vector{K}_\text{f}$ with zeros of size $M$
\FOR{each element in $\vector{K}_\text{f}$}
    \STATE $\vector{K}_\text{f} \leftarrow$ random number in $[0, 1)$
\ENDFOR
\STATE $\vector{K}_\text{f} \leftarrow (\vector{K}_\text{f} \times 2) // 1$
\STATE \RETURN $\vector{K}_\text{f}$
    \end{algorithmic}
\end{algorithm}
\end{figure}

\subsubsection{Flipping}\label{sec:bit-speech-enc}
In this subsection, we describe an encryption method for speech data using ``Flipping'', which uses a sign inversion as the secret key.
Algorithm~\ref{alg:bf} outlines the algorithm for encrypting speech data using Flipping.
\begin{enumerate}
\item The speech data~$\vector{X}$ is divided into blocks~$\vector{X}_{ij}$ of block size~$M$ according to Eq.~\eqref{eq:whole}.

\item To encrypt each block~$\vector{X}_{ij}$ using a secret key, $\vector{X}_{ij}$ is flattened to a one-dimensional vector~$\hat{\vector{X}}_{ij}$ according to Eq.~\eqref{eq:flatten}.

\item The secret key~$\vector{K}_\text{f}$ is generated. 
$\vector{K}_\text{f}$ is denoted as follows: 
\begin{equation}\label{eq:f}
\vector{K}_\text{f}=\{ k_\text{f}(1), k_\text{f}(2), \dots, k_\text{f}(k), \dots, k_\text{f}(N)\},
\end{equation}
where $k_\text{f}(k)\in \{0,1\}$, $P(X=k_\text{f}(k))= 0.5$, $1\leq k \leq N$, $Pr(X=p)$ represents the probability that $X$ takes the value $p$.

\item The matrix~$\vector{K}'_\text{f}$ used for encryption is generated as follows:
\begin{equation}\label{eq:key_f}
\vector{K}'_\text{f}(k) = 
\begin{cases}
-1 & (k_\text{f}(k)=1)\\
1  & (k_\text{f}(k)=0)
\end{cases}.
\end{equation}

\item The Hadamard product of $\hat{\vector{X}}_{ij}$ and $\vector{K}'_\text{f}$ is calculated to obtain the encrypted row vector~$\hat{\vector{X}}_{ij}^{(\boldsymbol{K}_\text{f})}$ as follows:
\begin{eqnarray}
\hat{\vector{X}}_{ij}^{(\boldsymbol{K}_\text{f})}   & =  &  
\vector{K}'_\text{f}\odot\hat{\vector{X}}_{ij} \nonumber\\
  & =  &  
\left[ \vector{K}'_\text{f}(1), \vector{K}'_\text{f}(2), \dots, \vector{K}'_\text{f}(N)  \right] \odot \left[ {x}_1, \dots, {x}_N \right]\nonumber\\
  & =  & 
\left[ 
{x}^{(k_\text{f})}_{1}, \dots, {x}^{(k_\text{f})}_{N}
\right] .
\label{eq:enc_q_f}
\end{eqnarray}

\item The one-dimensional vector~$\hat{\vector{X}}_{ij}^{(\boldsymbol{K}_\text{f})}$ is reshaped using the reshape function so that $\hat{\vector{X}}_{ij}^{(\boldsymbol{K}_\text{f})}$ equals the unencrypted block~$\vector{X}_{ij}$ according to Eq.~\eqref{eq:reshape}, and the encrypted block~$\vector{X}_{ij}^{(\boldsymbol{K}_\text{f})}$ is obtained.

\item All the blocks of the speech data~$\vector{X}$ that were divided in step $1$ are processed in steps $2$ - $6$ to obtain the encrypted speech data~$\vector{X}^{(\boldsymbol{K}_\text{f})}$.
\end{enumerate}

The steps from 1 to 7 demonstrate the transformation of plain speech data into encrypted speech data with the secret key $\vector{K}_\text{f}$.

\begin{figure}[tp]
\begin{algorithm}[H]
    \caption{ROM speech data encryption}
    \label{alg:rom}
    \begin{algorithmic}[1] 
    \REQUIRE $\vector{X}\text{: speech data}$, $M\text{: block size}$, $\vector{K}_\text{r}\text{: secret key}$
    \ENSURE $\vector{X}^{(\boldsymbol{K}_\text{r})}$: encrypted speech data
    \STATE $\vector{X}^{(\boldsymbol{K}_\text{r})} \leftarrow O$
    \STATE $f \leftarrow [F/M]$
    \STATE $t \leftarrow [T/M]$
    \FOR{$((i=0; i<f; i=i+M))$}
    \FOR{$((j=0; j<t; j=j+M))$}
    \STATE $\vector{X}_{ij} \leftarrow \vector{X}[i:i+M, j:j+M]$
    \STATE $\hat{\vector{X}}_{ij} \leftarrow \mathrm{flatten}(\vector{X}_{ij})$
    \STATE $\hat{\vector{X}}_{ij}^{(\boldsymbol{K}_\text{r})} \leftarrow \hat{\vector{X}}_{ij}^{(\boldsymbol{K}_\text{r})}\vector{K}_\text{r}$
    \STATE $\vector{X}_{ij}^{(\boldsymbol{K}_\text{r})} \leftarrow \mathrm{reshape}(\hat{\vector{X}}_{ij}^{(\boldsymbol{K}_\text{r})})$
    \STATE $\vector{X}^{(\boldsymbol{K}_\text{r})}[i:i+M, j:j+M] \leftarrow \vector{X}_{ij}^{(\boldsymbol{K}_\text{r})}$
    \ENDFOR
    \ENDFOR
    \RETURN $\vector{X}^{(\boldsymbol{K}_\text{r})}$
    \end{algorithmic}
\end{algorithm}
\end{figure}

\begin{figure}[tp]
\begin{algorithm}[H]
    \caption{Secret key generation in ROM~\cite{mezzadri2007generate}}
    \label{alg:scr-rom}
    \begin{algorithmic}[1] 
    \REQUIRE $M$: block size 
\ENSURE $\vector{K}_\text{r}$: secret key
\STATE $\vector{A} \leftarrow \text{random normal matrix of size } M \times M$
\STATE $(\vector{Q}, \vector{R}) \leftarrow \text{QR decomposition of } \vector{A}$
\FOR {$i = 1$ to $M$}
    \IF {$\vector{R}[i, i] < 0$}
        \STATE $Q[:,i] \leftarrow -Q[:,i]$
    \ENDIF
\ENDFOR
\STATE $\vector{K}_\text{r} \leftarrow Q$
\STATE \RETURN $\vector{K}_\text{r}$
    \end{algorithmic}
\end{algorithm}
\end{figure}

\subsubsection{Random orthogonal matrix}\label{sec:orth-speech-enc}
In this section, we describe an encryption method for speech data using ``ROM'', which uses a randomly generated orthogonal matrix as a secret key.
Algorithm~3 outlines the algorithm for encrypting speech data using ROM.

\begin{enumerate}
\item The speech data~$\vector{X}$ is divided into blocks~$\vector{X}_{ij}$ of block size~$M$ according to Eq.~\eqref{eq:whole}.

\item To encrypt each block~$\vector{X}_{ij}$ using a secret key, $\vector{X}_{ij}$ is flattened to a one-dimensional vector~$\hat{\vector{X}}_{ij}$ according to Eq.~\eqref{eq:flatten}.

\item The secret key~$\vector{K}_\text{r}$ is generated. 
$\vector{K}_\text{r}$ is denoted as follows: 
\begin{equation}\label{eq:key_r}
\vector{K}_\text{r}=
\begin{bmatrix}
k_{11}    &  \dots   &  k_{1N}\\
\vdots   &    \ddots   &       \vdots  \\
k_{N1}    &  \dots   &  k_{NN}\\
\end{bmatrix}
= \left[ \vector{k}_1, \vector{k}_2, \dots, \vector{k}_N \right].
\end{equation}

\item The matrix product of $\hat{\vector{X}}_{ij}$ and $\vector{K}_\text{r}$ is calculated to obtain the encrypted row vector~$\hat{\vector{X}}_{ij}^{(\boldsymbol{K}_\text{r})}$ as follows:
\begin{eqnarray}
\hat{\vector{X}}_{ij}^{(\boldsymbol{K}_\text{r})}   & =  &  
\hat{\vector{X}}_{ij}\vector{K}_\text{r} \nonumber\\
  & =  &  
\left[ {x}_1, \dots, {x}_N \right] 
\left[ \vector{k}_1, \dots, \vector{k}_N \right] \nonumber\\
  & =  & 
\left[ 
{x}^{(k_\text{r})}_{1}, \dots, {x}^{(k_\text{r})}_{N}
\right] .
\label{eq:enc_q}
\end{eqnarray}

\item The one-dimensional vector~$\hat{\vector{X}}_{ij}^{(\boldsymbol{K}_\text{r})}$ is reshaped using the reshape function so that $\hat{\vector{X}}_{ij}^{(\boldsymbol{K}_\text{r})}$ is equal to the unencrypted block~$\vector{X}_{ij}$ according to Eq.~\eqref{eq:reshape}, and the encrypted block~$\vector{X}_{ij}^{(\boldsymbol{K}_\text{r})}$ is obtained.

\item All the blocks of the speech data~$\vector{X}$ that were divided in step $1$ are processed in steps $2$ - $5$ to obtain the encrypted speech data~$\vector{X}^{(\boldsymbol{K}_\text{r})}$.
\end{enumerate}


\begin{figure}[t]
\centering
  \centering
    \includegraphics[keepaspectratio,width=11cm]{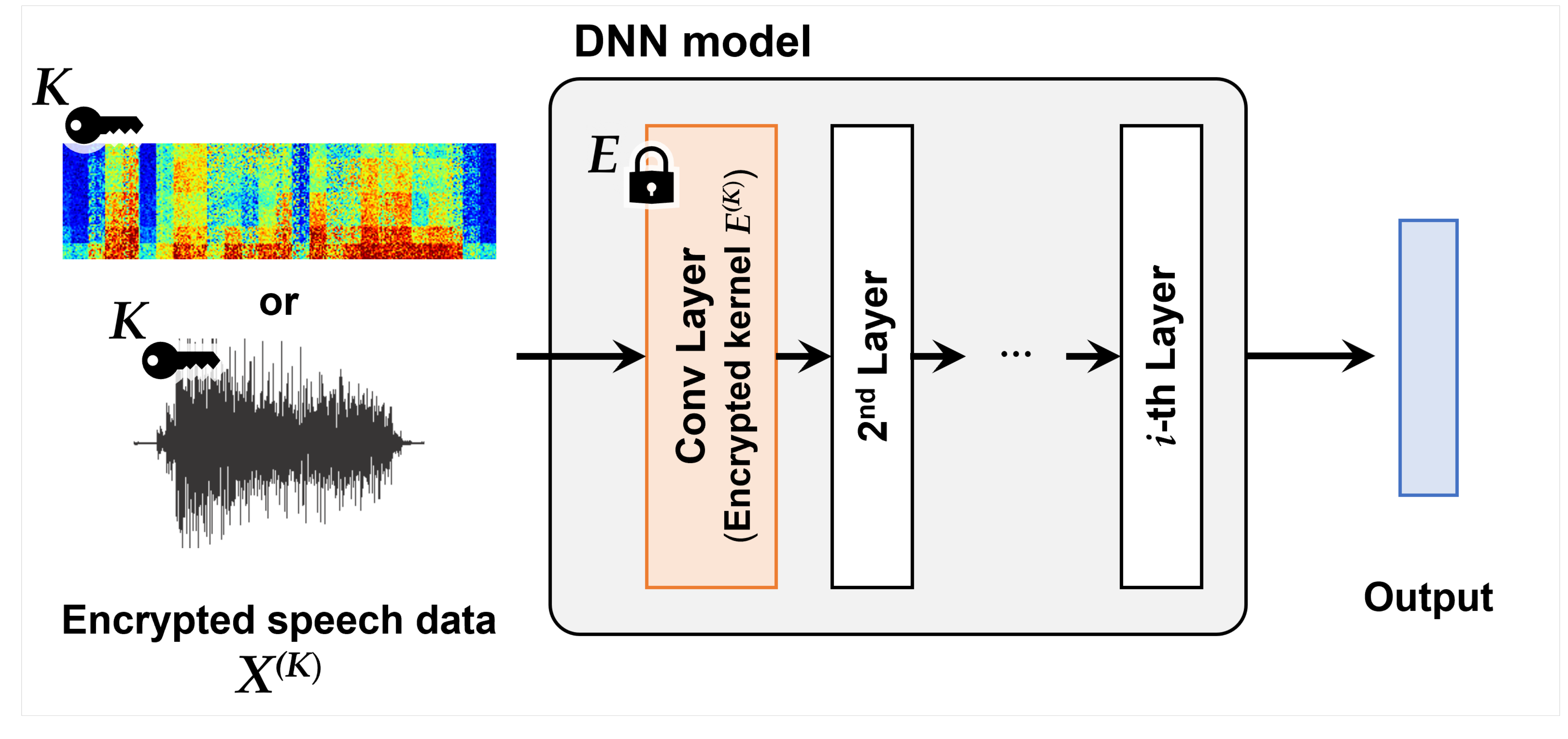}
  \caption{Encrypted models for accepting encrypted speech data}
  \label{figs:model-enc}
\end{figure}

\subsection{Model encryption}\label{sec:model-enc}
To input speech data encrypted using the procedure shown in subsection~\ref{sec:queryenc} directly into the trained model without decrypting it, it is required to transform a part of the trained model.
Therefore, we assume that the first layer of the CNN model is a convolutional layer, and that the kernel size and stride size of the first convolutional layer are equal.
This is because equal kernel and stride sizes allow the convolution process to be performed for each encrypted block.
Let $\vector{E}$ be the kernel of the first convolutional layer in which encryption is applied as shown in Fig.~\ref{figs:model-enc}, and let $P$ be the kernel size. Then, the kernel~$\vector{E}$ can be expressed as follows:
\begin{equation}\label{eq:kernel}
\vector{E}=
\begin{bmatrix}
e_{11}    &  e_{12}   &  \dots   &  e_{1b}\\
e_{21}    &  e_{22}   &  \dots   &  e_{2b}\\
\vdots   &   \vdots   &    &       \vdots  \\
e_{a1}    &  e_{a2}   &  \dots   &  e_{ab}\\
\end{bmatrix},
\end{equation}
where the size of the kernel~$\vector{E}$ can be expressed as $a\times b$.
When the speech data~$\vector{X}$ is one-dimensional data, $a=1$ and $b=P$, and when it is two-dimensional data, $a=P$ and $b=P$.
In addition, in this paper, the kernel size and the block size~$M$ used for encryption are assumed to be equal, i.e., $P=M$.
On the basis of these assumptions, when plain speech data~$\vector{X}$ is input to the convolution layer, a convolution operation is performed using the divided block~$\vector{X}_{ij}$ and the kernel~$\vector{E}$ as follows:
\begin{equation}\label{eq:conv}
z = \vector{X}_{ij} \cdot \vector{E} .
\end{equation}
We consider a scenario where the input data consists of encrypted speech data and aim to eliminate the effects of encryption without decrypting the encrypted speech data by using the kernel~$\vector{E}$.
The encryption procedures for the kernel~$\vector{E}$ for Shuffling, Flipping, and ROM are detailed in subsections \ref{sec:pix-kernel-enc}, \ref{sec:bit-kernel-enc}, and \ref{sec:orth-kernel-enc}, respectively.

\subsubsection{Shuffling}\label{sec:pix-kernel-enc} 
The encryption procedure of the model when using the secret key~$\vector{K}_\text{s}$ obtained from Shuffling will be explained.
First, the kernel~$\vector{E}$ is flattened to a one-dimensional row vector using the $\mathrm{flatten}$ function to obtain $\hat{\vector{E}}$.
Next, the matrix product of the permutation matrix~$\vector{K}'_\text{s}$ and the transposed $\hat{\vector{E}}$ is calculated to obtain the encrypted column vector~$\hat{\vector{E}}^{(\boldsymbol{K}_\text{s})}$ as follows:
\begin{equation}\label{eq:enc_kernel_s}
\hat{\vector{E}}^{(\boldsymbol{K}_\text{s})}=\vector{K}'_\text{s}\hat{\vector{E}}^t =
\begin{bmatrix}e^{(k_\text{s})}_{11}, e^{(k_\text{s})}_{12}, \cdots, e^{(k_\text{s})}_{ab} \end{bmatrix}^t.
\end{equation}
The column vector~$\hat{\vector{E}}^{(\boldsymbol{K}_\text{s})}$ is reshaped to be a matrix of the same size as the unencrypted kernel~$\vector{E}$ to obtain the encrypted kernel~$\vector{E}^{(\boldsymbol{K}_\text{s})}$ as follows:
\begin{eqnarray}
\vector{E}^{(\boldsymbol{K}_\text{s})}   & =  &  
\mathrm{reshape}(\hat{\vector{E}}^{(\boldsymbol{K}_\text{s})}) \nonumber \\
  & =  & 
\begin{bmatrix}
e^{(k_\text{s})}_{11}    &  e^{(k_\text{s})}_{12}   &  \dots   &  e^{(k_\text{s})}_{1b}\\
e^{(k_\text{s})}_{21}    &  e^{(k_\text{s})}_{22}   &  \dots   &  e^{(k_\text{s})}_{2b}\\
\vdots   &   \vdots   &    &       \vdots  \\
e^{(k_\text{s})}_{a1}    &  e^{(k_\text{s})}_{a2}   &  \dots   &  e^{(k_\text{s})}_{ab}
\end{bmatrix}.
\label{eq:reshape_kernel}
\end{eqnarray}
When calculating convolution with the encrypted speech data~$\vector{X}^{(\boldsymbol{K}_\text{s})}$ and the encrypted kernel~$\vector{E}^{(\boldsymbol{K}_\text{s})}$, the computation for each encrypted block~$\vector{X}_{ij}^{(\boldsymbol{K}_\text{s})}$ is as follows:
\begin{eqnarray}\label{eq:enc-conv_s}
z^{(\boldsymbol{K}_\text{s})} &=& \vector{X}_{ij}^{(\boldsymbol{K}_\text{s})} \cdot \vector{E}^{(\boldsymbol{K}_\text{s})} = \hat{\vector{X}}_{ij}^{(\boldsymbol{K}_\text{s})} \hat{\vector{E}}^{(\boldsymbol{K}_\text{s})}\nonumber\\
&=& \hat{\vector{X}}_{ij} \vector{K}'_\text{s} \vector{K}_\text{s}^{'t} \hat{\vector{E}}^{t} = \vector{X}_{ij} \cdot \vector{E} = z.
\end{eqnarray}
Since the permutation matrix is a kind of orthogonal matrix, the product of $\vector{K}'_\text{s}$ and $\vector{K}^{'t}_\text{s}$ is a unit matrix according to the property of the orthogonal matrix.
By inputting speech data encrypted with the same secret key used to encrypt the model into the encrypted model, the internal computations produce identical results as if no encryption had been performed. 
Therefore, it is possible to input encrypted speech data into the model without decryption, allowing for the correct utilization of the model while preserving the privacy of the speech data.
On the other hand, when encrypted speech data, encrypted using a different secret key from that used for the model encryption, is input into the encrypted model, the results differ from those obtained without encryption. Therefore, it is hard to use the model correctly without prior knowledge of the secret key.
\textcolor{black}{In this framework, since the encryption process is applied to the original speech data after it has been recorded, the presence of noise in the original speech data does not affect the encryption.
As shown in Eq.~\eqref{eq:enc-conv_s}, when the correct key is used, the impact of the secret key is canceled out during the inner product calculation. While the performance of the original model may degrade if it is not robust to noise, this is due to the inherent characteristics of the model and not due to the proposed encryption method.
Our method does not interfere with the inner product computation the model performs when the correct key is used.}

\subsubsection{Flipping}\label{sec:bit-kernel-enc}
The encryption procedure of the model when using the secret key~$\vector{K}_\text{f}$ obtained from Flipping will be explained.
First, the kernel~$\vector{E}$ is flattened to a one-dimensional row vector using the $\mathrm{flatten}$ function to obtain $\hat{\vector{E}}$.
Next, the Hadamard product of $\vector{K}'_\text{f}$ and $\hat{\vector{E}}$ is calculated to obtain the encrypted column vector~$\hat{\vector{E}}^{(\boldsymbol{K}_\text{f})}$ as follows:
\begin{equation}\label{eq:enc_kernel_f}
\hat{\vector{E}}^{(\boldsymbol{K}_\text{f})}=\vector{K}_\text{f}^{'t}\odot\hat{\vector{E}}^t =
\begin{bmatrix}e^{(k_\text{f})}_{11}, e^{(k_\text{f})}_{12}, \cdots, e^{(k_\text{f})}_{ab} \end{bmatrix}^t.
\end{equation}
The column vector~$\hat{\vector{E}}^{(\boldsymbol{K}_\text{f})}$ is reshaped to be a matrix of the same size as the unencrypted kernel~$\vector{E}$ to obtain the encrypted kernel~$\vector{E}^{(\boldsymbol{K}_\text{f})}$ according to Eq.~\eqref{eq:reshape_kernel}.
When calculating convolution with the encrypted speech data~$\vector{X}^{(\boldsymbol{K}_\text{f})}$ and the encrypted kernel~$\vector{E}^{(\boldsymbol{K}_\text{f})}$, the computation for each encrypted block~$\vector{X}_{ij}^{(\boldsymbol{K}_\text{f})}$ is as follows:
\begin{eqnarray}\label{eq:enc-conv_f}
z^{(\boldsymbol{K}_\text{f})} &=& \vector{X}_{ij}^{(\boldsymbol{K}_\text{f})} \cdot \vector{E}^{(\boldsymbol{K}_\text{f})} = \hat{\vector{X}}_{ij}^{(\boldsymbol{K}_\text{f})} \hat{\vector{E}}^{(\boldsymbol{K}_\text{f})} = 
(\vector{K}'_\text{f} \odot \vector{X}_{ij}^{(\boldsymbol{K}_\text{f})})(\vector{K}_\text{f}^{'t} \odot \vector{\hat{E}}^t)\nonumber\\
&=& 
\begin{pmatrix}
[\vector{K}'_\text{f}(1) \dots \vector{K}'_\text{f}(N)]\odot
[x_1 \dots x_N]
\end{pmatrix}
\begin{pmatrix}
\begin{bmatrix}
\vector{K}'_\text{f}(1)\\ \vdots\\ \vector{K}'_\text{f}(N)
\end{bmatrix}\odot
\begin{bmatrix}
e_{11}\\ \vdots\\ e_{ab}
\end{bmatrix}
\end{pmatrix}\nonumber\\
&=& 
[\vector{K}'_\text{f}(1) x_1 \dots \vector{K}'_\text{f}(N) x_N]
\begin{bmatrix}
\vector{K}'_\text{f}(1) e_{11}\\ \vdots\\ \vector{K}'_\text{f}(N) e_{ab}
\end{bmatrix} 
= \vector{X}_{ij} \cdot \vector{E} = z.
\end{eqnarray}
Since $\vector{K}'_\text{f}$ is a matrix consisting of $-1$ or $1$, we can obtain completely the same results before and after using the proposed method by inputting speech data encrypted with the secret key~$\vector{K}_\text{f}$ to the model encrypted with the same secret key~$\vector{K}_\text{f}$.
Therefore, as well as with Shuffling, it is possible to input the encrypted speech data into the model without decrypting it.

\subsubsection{Random orthogonal matrix}\label{sec:orth-kernel-enc}
The encryption procedure of the model when using the secret key~$\vector{K}_\text{r}$ obtained from ROM will be explained.
First, the kernel~$\vector{E}$ is flattened to a one-dimensional row vector using the $\mathrm{flatten}$ function to obtain $\hat{\vector{E}}$.
Next, the matrix product of $\vector{K}_\text{r}^t$ and $\hat{\vector{E}}^t$ is calculated to obtain the encrypted column vector~$\hat{\vector{E}}^{(\boldsymbol{K}_\text{r})}$ as follows:
\begin{equation}\label{eq:enc_kernel_r}
\hat{\vector{E}}^{(\boldsymbol{K})}=\vector{K}_\text{r}^t\hat{\vector{E}}^t =
\begin{bmatrix}e^{(k)}_{11}, e^{(k)}_{12}, \cdots, e^{(k)}_{ab} \end{bmatrix}^t.
\end{equation}
The column vector~$\hat{\vector{E}}^{(\boldsymbol{K}_\text{r})}$ is reshaped to be a matrix of the same size as the unencrypted kernel~$\vector{E}$ to obtain the encrypted kernel~$\vector{E}^{(\boldsymbol{K}_\text{r})}$ according to Eq.~\eqref{eq:reshape_kernel}.
When calculating convolution with the encrypted speech data~$\vector{X}^{(\boldsymbol{K}_\text{r})}$ and the encrypted kernel~$\vector{E}^{(\boldsymbol{K}_\text{r})}$, the computation for each encrypted block~$\vector{X}_{ij}^{(\boldsymbol{K}_\text{r})}$ is as follows:
\begin{eqnarray}\label{eq:enc-conv}
z^{(\boldsymbol{K})} &=& \vector{X}_{ij}^{(\boldsymbol{K})} \cdot \vector{E}^{(\boldsymbol{K})} = \hat{\vector{X}}_{ij}^{(\boldsymbol{K})} \hat{\vector{E}}^{(\boldsymbol{K})}\\\nonumber
&=& \hat{\vector{X}}_{ij} \vector{K}_\text{r} \vector{K}_\text{r}^{t} \hat{\vector{E}}^{t} = \vector{X}_{ij} \cdot \vector{E} = z.\nonumber
\end{eqnarray}
The matrix product of $\vector{K}_\text{r}$ and $\vector{K}^{t}_\text{r}$ is a unit matrix according to the property of the orthogonal matrix.
By inputting speech data encrypted with a secret key~$\vector{K}_\text{r}$ into a model encrypted with the same secret key~$\vector{K}_\text{r}$, completely the same results can be obtained before and after using the proposed method.
Therefore, as well as with Shuffling and Flipping, it is possible to input the encrypted speech data into the model without decoding it.

\subsection{Key space of secret key}
In this section, we discuss the key space size for the secret keys utilized in Shuffling, Flipping, and ROM.
Concerning Shuffling, as depicted in Eq.~\eqref{eq:s}, the secret key~$\vector{K}_\text{s}$ is a matrix for rearranging the indices of elements within each block~$\vector{X}_{ij}$ in any order, e.g., $\vector{K}_\text{s}=[3, 1, 2]$ when the input data~$\vector{X}$ is one-dimensional and $M=3$.
Therefore, when the speech data~$\vector{X}$ is one-dimensional, the secret key~$\vector{K}_\text{s}$ can have $M!$ possible patterns, and when the speech data~$\vector{X}$ is two-dimensional, it can have $(M\times M)!$ possible patterns.
Concerning Flipping, as depicted in Eq.~\eqref{eq:f}, the secret key~$\vector{K}_\text{f}$ is a bit sequence consisting of $0$ or $1$, e.g., $\vector{K}_\text{f}=[0,0,1]$ when the input data is one-dimensional and $M=3$.
Therefore, the secret key~$\vector{K}_\text{f}$ can only be used in $2^{M}$ ways when the speech data~$\vector{X}$ is one-dimensional and in $2^{M \times M}$ ways when the speech data~$\vector{X}$ is two-dimensional.
Since each bit in $K_f$ is generated independently with a probability of 0.5, all $2^M$ or $2^{M\times M}$ patterns occur with equal probability.
Concerning ROM, as depicted in Eq.~\eqref{eq:key_r}, the secret key~$\vector{K}_\text{r}$ is an orthogonal matrix and is composed of randomly generated real-valued elements, including negative values.
As an example of the secret key~$\vector{K}_\text{r}$, the matrix for the case where the input data is one-dimensional and $M=3$ is shown below:
\begin{equation}\label{eq:okey}
\vector{K}_\text{r}=
\begin{bmatrix}
0.9898  & -0.0661  & -0.1264\\
0.1309  &  0.7732  &  0.6205\\
0.0568  & -0.6307  &  0.7740\\
\end{bmatrix}.
\end{equation}
As shown in Eq.~\eqref{eq:okey}, $\vector{K}_\text{r}$ is a matrix of $M\times M$ when the speech data~$\vector{X}$ is one-dimensional and $M^2\times M^2$ when $\vector{X}$ is two-dimensional, allowing for the generation of many patterns of secret keys.
Focusing on periods of silence in the speech data, there is a risk that the secret key can be easily estimated by a third party if the keyspace is small.
Therefore, it is better to have a large key space for the secret key to make the prediction of the secret key more difficult.
Despite the limitation that the secret key must be an orthogonal matrix, ROM has the advantage of the key space of the secret key being larger than that of Shuffling and Flipping, making the prediction of the secret key much more difficult.

\subsection{Key generation procedure}
This section provides the concrete procedure to generate a secret key pair. 

For Shuffling, the Python library \verb|torch.randperm(n=M)| is employed to generate the secret key $\vector{K}_\text{s}$. Then, its transpose matrix $\vector{K}_\text{s}^t$ is calculated with \verb|torch.t|.

For ROM, the Python library \verb|scipy.stats.ortho_group.|\verb|rvs(dim=M)| is employed to generate the secret key $\vector{K}_\text{r}$. Then, its transpose matrix $\vector{K}_\text{r}^t$ is calculated with \verb|torch.t|.

For the evaluation, many secret key pairs are generated by changing the seed value, and the keys of different pairs are regarded as incorrect.


\section{Experiment}\label{seq:experiment}
\subsection{Evaluation of privacy-preserving performance}
\subsubsection{Experimental conditions}
In this experiment, we evaluated the privacy-preserving performance of the proposed methods using ASV, ASR, and ASC tasks.


ASV is a technology used to verify the identity of a speaker by analyzing their speech characteristics. The task of ASV involves determining whether a claimed speaker matches the true identity by comparing speech samples. 
Within the privacy-preserving scenario of the ASV task, encryption is used with the aim of ensuring that authentication is only successful when the correct secret key is used, while performance significantly deteriorates when an incorrect key is utilized. 
The ASV experiment is assumed to assess privacy-preserving performance against the speaker's identity. 
For the ASV system, we used an x-vector-based ASV system~\cite{xvector} with a self-supervised-learning (SSL) based front-end model~\cite{yang21c_interspeech, NEURIPS2020_92d1e1eb}.
For the ASV task, we trained a HuBERT model~\cite{hubert} with the LibriSpeech corpus~\cite{libri} following the Fairseq recipe~\cite{ott2019fairseq}.
The HuBERT model is used as an SSL-based front-end model, and it is regarded as one of the state-of-the-art systems.
The input features for the HuBERT model are waveforms.
The structure and hyperparameters of the HuBERT model were the same as those of HuBERT Base~\cite{hubert}, except that the stride size~$P$ of the first convolutional layer was changed to ten.
The speech expression outputted from the HuBERT model was inputted to an x-vector-based embedding network. 
This network was trained with the VoxCeleb1 corpus~\cite{voxceleb}, using the same hyperparameters as in~\cite{yang21c_interspeech}. 
For the ASV evaluation, we used the VoxCeleb1 test set, and the input waveforms were encrypted by the proposed encryption methods. 
The block size~$M$ for the encryption methods was set to $10$.
Equal error rate (EER) was used as the evaluation metric.

ASR is the process of transcribing speech content into text.
Within the privacy-preserving scenario of the ASR task, encryption is performed with the aim of ensuring that recognition is only successful when the correct secret key is used, while the performance of the speech recognition deteriorates significantly when an incorrect key is utilized.
The ASR experiment is assumed to assess privacy-preserving performance against the speech content. 
For the ASR task, we trained a transformer model with the LibriSpeech corpus following the ESPnet2 recipe~\cite{Watanabe2018}.
The transformer architecture and hyperparameters were the same as in~\cite{ESPnetASR}, except for the input feature and the stride size of the first convolutional layer.
The acoustic features are input as two-dimensional log-mel spectrogram features by arranging the 80-dimensional log-mel filterbank features extracted for each frame.
The stride size~$P$ of the first convolutional layer was set to three to use the proposed methods.
For the ASR evaluation, we used the LibriSpeech test clean subset, and the input log-mel spectrogram features were encrypted by the proposed methods. 
The block size~$M$ for the encryption was set to three to match the kernel size~$P$.
Word error rate~(WER) was used as an evaluation metric.
We also evaluated how block size influences the performance of the proposed methods 
in concealing the speech content within the encrypted speech.
Under $M=5,10,20,128$, speech waveforms encrypted using the proposed methods were input to the plain pre-trained speech recognition model published in \cite{ESPnetASR}.

ASC is the task of categorizing audio recordings depending on the type or category of the surrounding environment in which they were captured~\cite{mesaros2018acoustic}. 
Unlike ASR, ASC focuses on classifying the acoustic event of various environments. 
ASC systems analyze features extracted from audio signals, such as spectrograms, and use machine learning algorithms to classify the audio into predefined categories or classes.
The ASC experiment is assumed to assess privacy-preserving performance against an acoustic event.
For the ASC task, we used the ConvMixer~\cite{trockman2022patches} model trained on the SINS~\cite{dekkers2017sins} dataset.
We used only the SINS data labeled Absence, Cooking, Dishwashing, Eating, Other, Vacuumcleaner, Watching TV, Working, Calling, and Visit, and data labeled Calling and Visit were combined into a single class and labeled Social Activity.
The input features for the ConvMixer model are spectrograms, which are two-dimensional speech data.
The data were clipped so that the length of each sample was four seconds, and spectrograms were generated from the clipped data.
The structure and hyperparameters of the ConvMixer model were the same as those of ConvMixer-768/32~\cite{trockman2022patches}, except that the stride size~$P$ and the kernel size of the first convolutional layer was changed to eight.
For the ASC evaluation, we used the test set of the SINS dataset, and input spectrograms were encrypted by the proposed methods. 
The block size~$M$ for the encryption methods was set to eight.
Accuracy, which indicates the percentage of correct classifications, was used as an index to evaluate the classification results.

\begin{table}[t]
\caption{EER$(\%)$ in encryption scenario ($M=10$) for ASV.}
\label{tab:asv-result}
\centering
\begin{tabular}{c|ccc}
\hline
\begin{tabular}[c]{@{}c@{}}\vspace{-1cm}Model encryption\end{tabular} & \multicolumn{3}{c}{Query encryption}                                              \\\cline{2-4}
                                                                   & Plain & \begin{tabular}[c]{@{}c@{}}Correct \\[-1mm]key\end{tabular} & \begin{tabular}[c]{@{}c@{}}Inorrect \\[-1mm]key\end{tabular}\\\hline\hline
Plain & 7.91  & -    & - \\
Shuffling & 36.7 & 7.91 & 33.3 \\
Flipping &  37.2 & 7.91 & 34.1 \\
ROM & 35.3  & 7.91 &  35.1 \\\hline
\end{tabular}
\end{table}

\subsubsection{Experimental results}
Tables~\ref{tab:asv-result}, \ref{tab:asr-result}, and \ref{tab:asc-result} show the results of the ASV, ASR and ASC experiments, respectively.
In these experiments, the plain model, i.e., no encryption, and the models encrypted by Shuffling, Flipping and ROM were used.
``Correct key'' refers to a situation where the encryption key used for the model matches the encryption key used for the query.
``Incorrect key'' refers to a situation where the encryption key used for the model does not match the encryption key used for the query.
The results of ``Incorrect key'' are based on the average values obtained when using five different incorrect secret keys.
``Plain'' in the context of query encryption can be regarded as a form of an incorrect key, indicating that encryption has not been performed.

Table~\ref{tab:asv-result} shows the results of the ASV experiments. 
The EERs for ``Correct key'' were completely the same as those of the plain model.
In the ``Incorrect key'' and ``Plain'' cases, the EERs were higher than those for ``Correct key.''
These results show that only authorized users who know the correct secret key can correctly use the encrypted model in the ASV task by using the proposed methods.
Furthermore, these results show that there is not much difference in the trend of the results between the methods, and that all of the methods succeed in concealing the speaker identity.

\begin{table}[t]
\caption{WER$(\%)$ in encryption scenario ($M=3$) for ASR.}
\label{tab:asr-result}
\centering
\begin{tabular}{c|ccc}
\hline
\begin{tabular}[c]{@{}c@{}}\vspace{-1cm}Model encryption\end{tabular} & \multicolumn{3}{c}{Query encryption}                                              \\\cline{2-4}
                                                                   & Plain & \begin{tabular}[c]{@{}c@{}}Correct \\[-1mm]key\end{tabular} & \begin{tabular}[c]{@{}c@{}}Incorrect \\[-1mm]key\end{tabular}\\\hline\hline
Plain& 4.4  & -    & - \\
Shuffling & 10.9  & 4.4  & 14.18 \\
Flipping &  98.1   & 4.4   & 98.26 \\
ROM & 99.7 & 4.4 & 97.62 \\\hline
\end{tabular}
\end{table}

Table~\ref{tab:asr-result} shows the results of the ASR experiments.
The WERs of the encrypted models for ``Correct key'' were completely the same as those of the plain model.
In the ``Incorrect key'' and ``Plain'' case, the WERs were higher than those of the ``Correct key'' case, especially in the Flipping and ROM case.
Therefore, in the ``Incorrect key'' and ``Plain'' case, the encrypted models hardly extracted the speech content, so the encryption by the proposed methods was highly anonymous.
As well as with the ASV results, these results also show that only authorized users who know the correct key can correctly use the encrypted model in the ASR task within the proposed methods.

Table~\ref{tab:asc-result} shows the results of the ASR experiments.
The accuracies of the encrypted models for ``Correct key'' were completely the same as those of the plain model.
In the ``Incorrect key'' case, the accuracies were higher than those of the ``Correct key'' case.
In the ``Plain'' scenario, the accuracy was also significantly increased since it can be regarded as one of the ``Incorrect key'' cases.
These results also show that only authorized users who know the correct key can correctly use the encrypted model in the ASC task by using the proposed methods.
On the basis of the privacy-preserving performance evaluation experiments described above, it is confirmed that the proposed methods prevent unauthorized users who do not know the secret key used to encrypt the model from using the model with high performance. 

Encrypted speech can also be obtained as severely noisy speech, so speech waveforms encrypted with Shuffling, Flipping, and ROM were input to the ASR model to investigate the encryption robustness of the proposed methods and the plain model,
and the results are shown in Tab.~\ref{tab:asr-result1}.
The WERs for the encrypted speech were higher in all conditions than the WER for the unencrypted speech, $2.7\%$, and we can see that for all encryption methods, the WER increased as the block size~$M$ increased.
Furthermore, when $M$ was small, for example $M=5,10$, the WER was higher when the waveform was encrypted by Flipping or ROM than when it was encrypted by Shuffling, indicating that the privacy-preserving performances of Flipping or ROM were better.
In particular, for encryption using ROM, the key space is sufficiently large even when $M$ is small, making it difficult to predict the secret key.

\begin{table}[t]
\caption{Accuracy (\%) in encryption scenario ($M=8$) for ASC.}
\label{tab:asc-result}
\centering
\begin{tabular}{c|ccc}
\hline
\begin{tabular}[c]{@{}c@{}}\vspace{-1cm}Model encryption\end{tabular} & \multicolumn{3}{c}{Query encryption}                                              \\\cline{2-4}
                                                                   & Plain & \begin{tabular}[c]{@{}c@{}}Correct \\[-1mm]key\end{tabular} & \begin{tabular}[c]{@{}c@{}}Incorrect \\[-1mm]key\end{tabular}\\\hline\hline
Plain& 85.4 & -    & - \\
Shuffling & 64.6 & 85.4 & 60.5 \\
Flipping & 1.97 & 85.4 & 1.99 \\
ROM & 1.97 & 85.4 & 1.99 \\\hline
\end{tabular}
\end{table}

\begin{table}[t]
\caption{Comparison of WER $(\%)$ of ASR model on LibriSpeech corpus (test clean subsets) encrypted using Shuffling, Flipping, and ROM. WER for unencrypted query input to plain model is $2.7\%$.}
\label{tab:asr-result1}
\centering
\begin{tabular}{c|ccc}
\hline
$M$                                                & Shuffling & Flipping & \begin{tabular}[c]{@{}c@{}}ROM\end{tabular}\\ \hline\hline
5    & 13.9 & 40.5 & 28.9      \\
10   & 30.6 & 68.1 & 57.6    \\
20   & 63.9 & 82.9 & 85.4        \\
128  & 94.9 & 95.0  & 94.8\\\hline       
\end{tabular}
\end{table}

\begin{figure}[t]
  \begin{tabular}{ccc}
      \hspace{-2mm}\begin{minipage}[b]{0.3\columnwidth}
    \centering
    \includegraphics[keepaspectratio,width=\columnwidth]{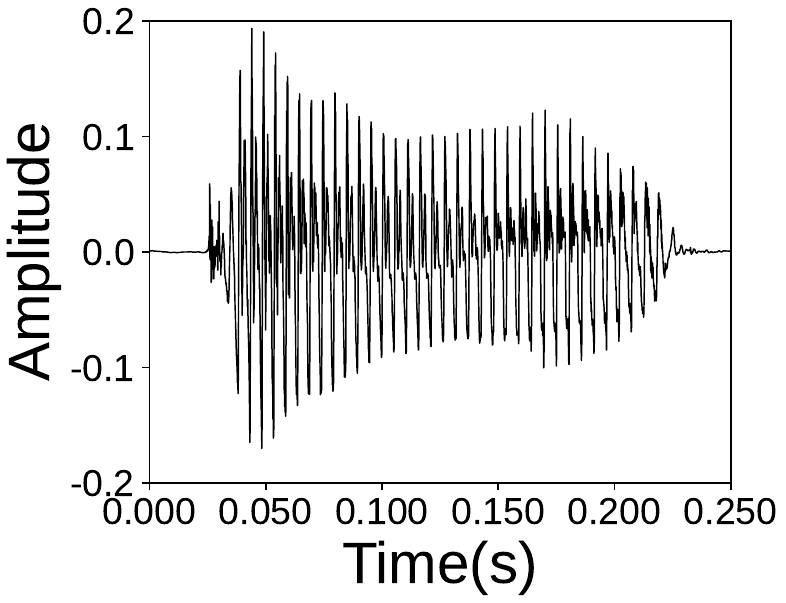}
    \subcaption{Original waveform}\vspace{0.8em}
    \label{pix-ori-wav}
  \end{minipage} &
      \hspace{-2mm}\begin{minipage}[b]{0.3\columnwidth}
    \centering
    \includegraphics[keepaspectratio, width=\columnwidth]{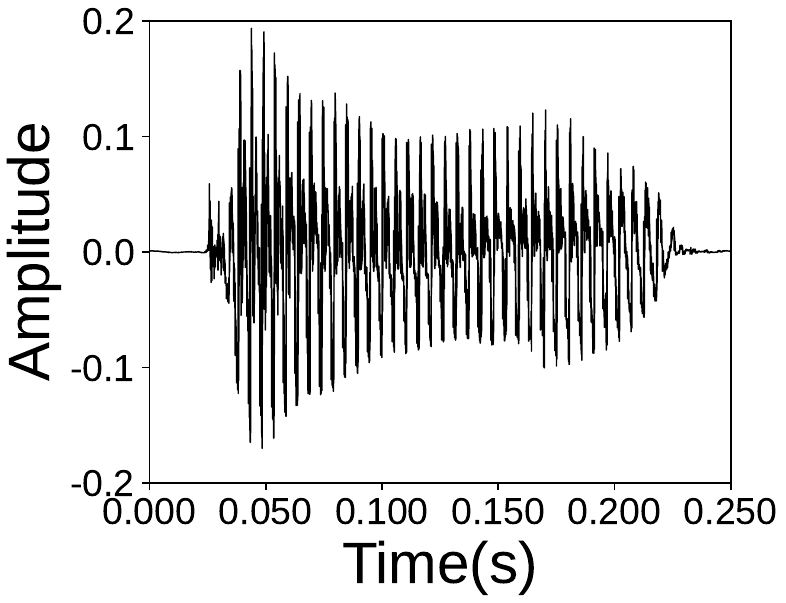}
    \subcaption{Encrypted waveform ($M=10$) }
    \label{pix10-wav}
  \end{minipage}&
  \hspace{-2mm}\begin{minipage}[b]{0.3\columnwidth}
    \centering
    \includegraphics[keepaspectratio, width=\columnwidth]{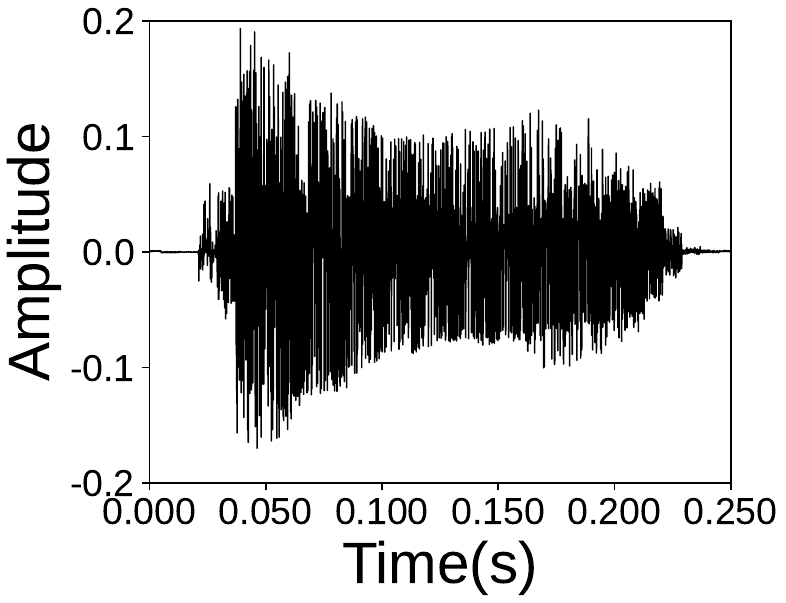}
    \subcaption{Encrypted waveform ($M=128$) }
    \label{pix128-wav}
  \end{minipage}\\

  \hspace{-2mm}\begin{minipage}[b]{0.3\columnwidth}
    \centering
    \includegraphics[keepaspectratio,width=\columnwidth]{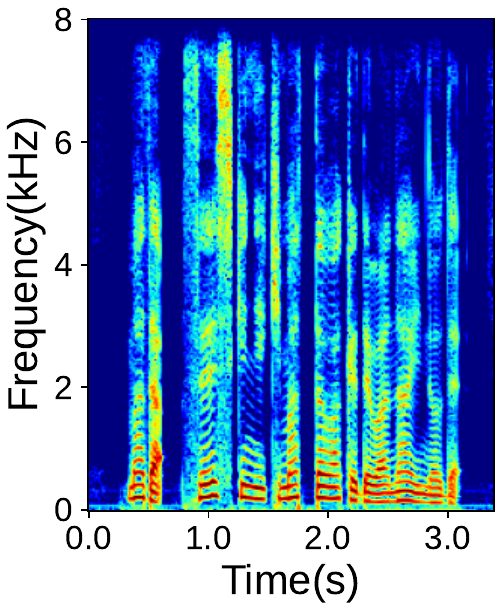}
    \subcaption{Original spectrogram}\vspace{0.8em}
    \label{pix-ori-wav-spe}
  \end{minipage} &
      \hspace{-2mm}\begin{minipage}[b]{0.3\columnwidth}
    \centering
    \includegraphics[keepaspectratio, width=\columnwidth]{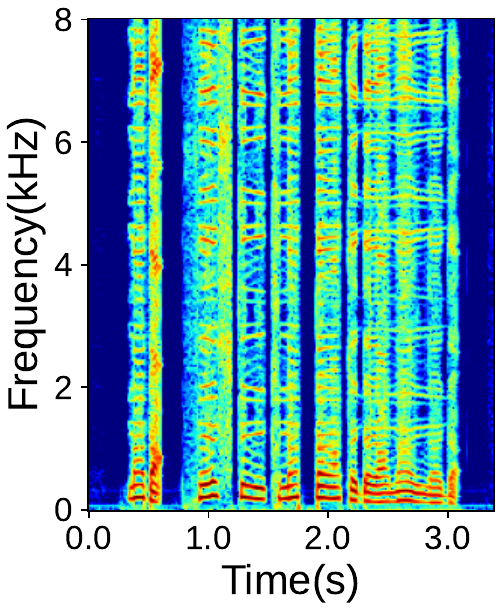}
    \subcaption{Encrypted spectrogram ($M=10$)}
    \label{pix10-wav-spe}
  \end{minipage}&
  \hspace{-2mm}\begin{minipage}[b]{0.3\columnwidth}
    \centering
    \includegraphics[keepaspectratio, width=\columnwidth]{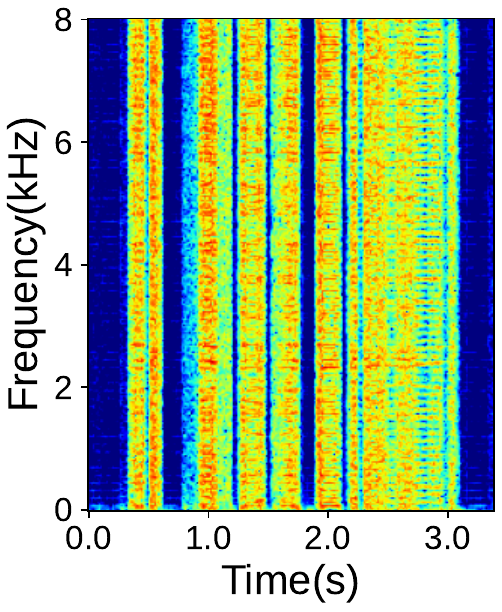}
    \subcaption{Encrypted spectrogram ($M=128$)}
    \label{pix128-wav-spe}
  \end{minipage}\\
  
  \end{tabular}
     \caption{Examples of waveform encrypted by Shuffling}
     \label{pix-wav-spe}
  \end{figure}

\begin{figure}[t]
\begin{tabular}{c|c}
  \begin{tabular}{cc}
      \hspace{-2mm}\begin{minipage}[b]{0.2\columnwidth}
    \centering
    \includegraphics[keepaspectratio, width=\columnwidth]{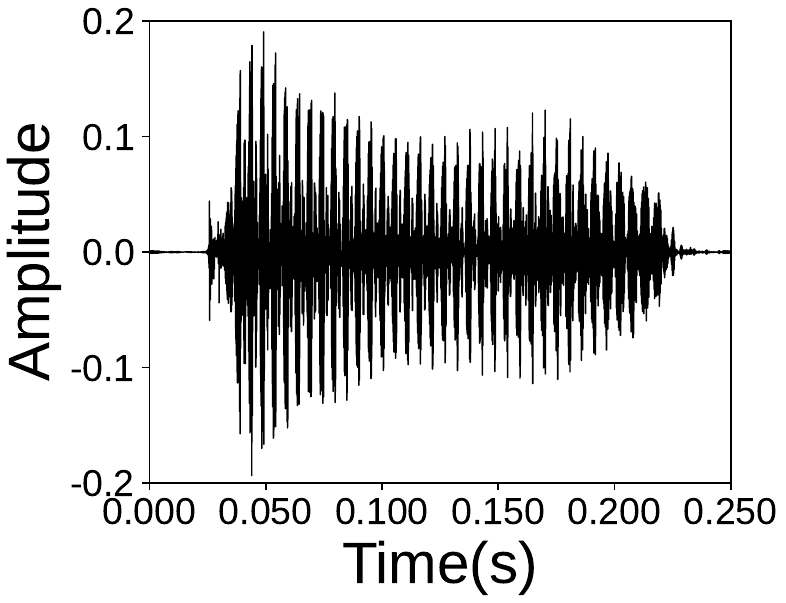}
    \subcaption{Encrypted waveform\\ ($M=10$) }
    \label{bit10-wav}
  \end{minipage}&
  \hspace{-2mm}\begin{minipage}[b]{0.2\columnwidth}
    \centering
    \includegraphics[keepaspectratio, width=\columnwidth]{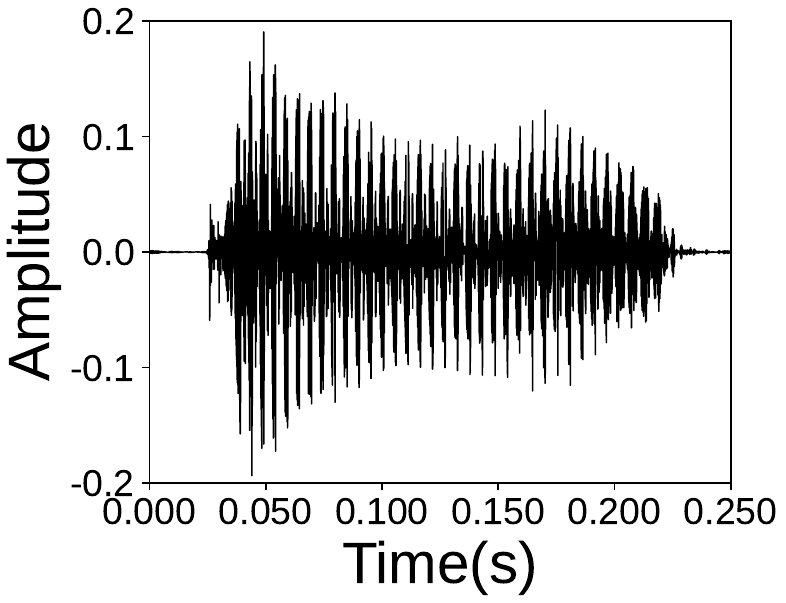}
    \subcaption{Encrypted waveform\\ ($M=128$) }
    \label{bit128-wav}
  \end{minipage}\\

      \hspace{-2mm}\begin{minipage}[b]{0.2\columnwidth}
    \centering
    \includegraphics[keepaspectratio, width=\columnwidth]{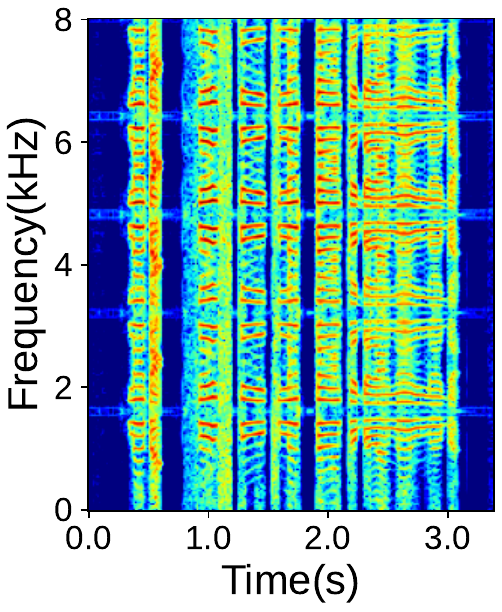}
    \subcaption{Encrypted spectrogram ($M=10$)}
    \label{bit10-wav-spe}
  \end{minipage}&
  \hspace{-2mm}\begin{minipage}[b]{0.2\columnwidth}
    \centering
    \includegraphics[keepaspectratio, width=\columnwidth]{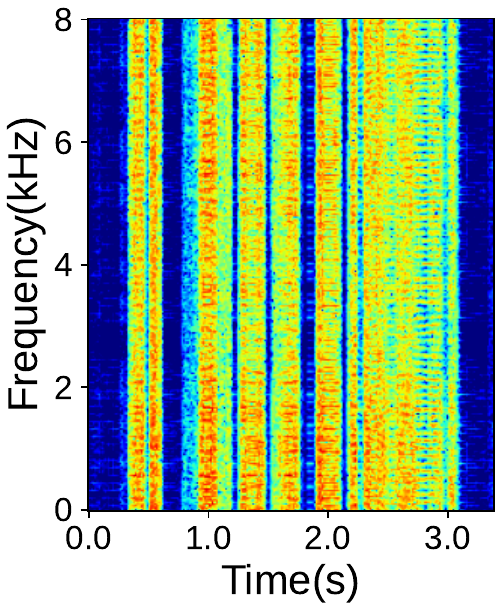}
    \subcaption{Encrypted spectrogram ($M=128$)}
    \label{bit128-wav-spe}
  \end{minipage}\\
  \multicolumn{2}{c}{\footnotesize \textbf{Flipping}}\\
  \end{tabular}
  &

\begin{tabular}{cc}
\centering
      \hspace{-2mm}\begin{minipage}[b]{0.2\columnwidth}
    \centering
    \includegraphics[keepaspectratio, width=\columnwidth]{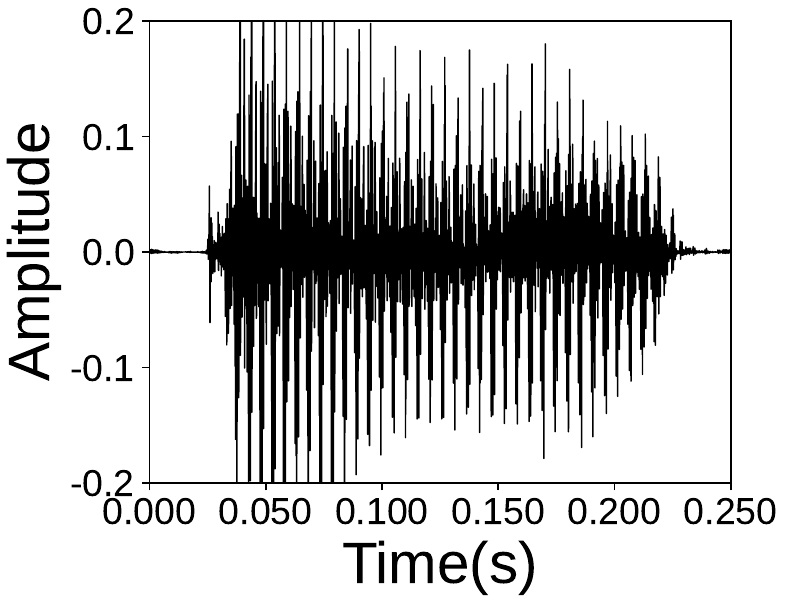}
    \subcaption{Encrypted waveform\\ ($M=10$) }
    \label{orth10-wav}
  \end{minipage}&
  \hspace{-2mm}\begin{minipage}[b]{0.2\columnwidth}
    \centering
    \includegraphics[keepaspectratio, width=\columnwidth]{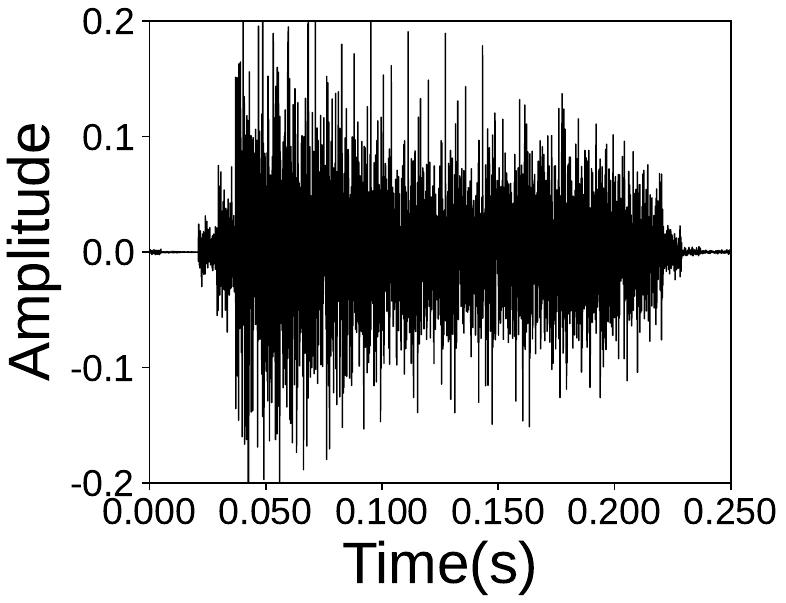}
    \subcaption{Encrypted waveform\\ ($M=128$) }
    \label{orth128-wav}
  \end{minipage}\\

      \hspace{-2mm}\begin{minipage}[b]{0.2\columnwidth}
    \centering
    \includegraphics[keepaspectratio, width=\columnwidth]{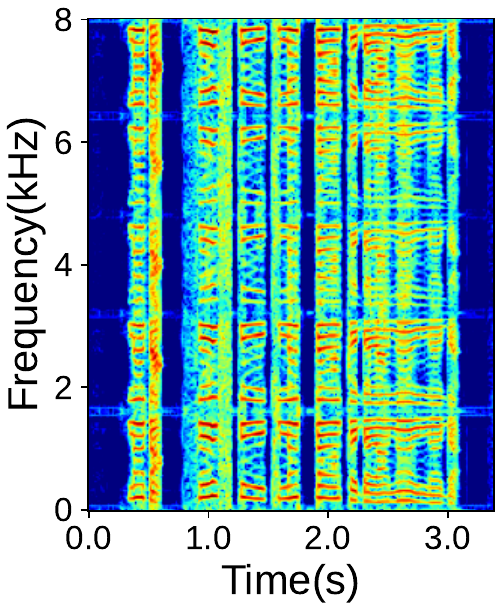}
    \subcaption{Encrypted spectrogram ($M=10$)}
    \label{orth10-wav-spe}
  \end{minipage}&
  \hspace{-2mm}\begin{minipage}[b]{0.2\columnwidth}
    \centering
    \includegraphics[keepaspectratio, width=\columnwidth]{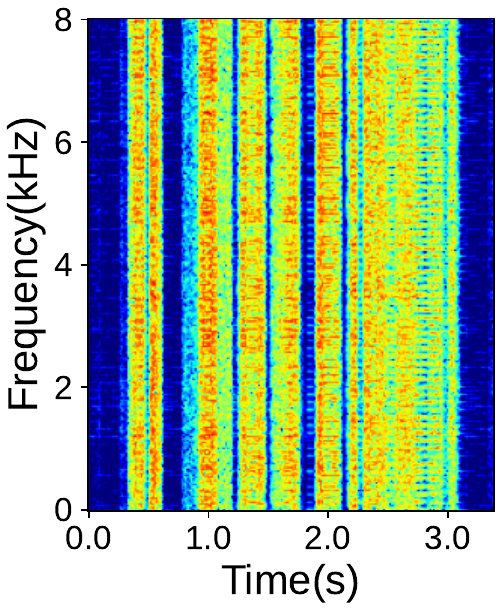}
    \subcaption{Encrypted spectrogram ($M=128$)}
    \label{orth128-wav-spe}
  \end{minipage}\\
  \multicolumn{2}{c}{\footnotesize \textbf{ROM}}\\
  \end{tabular}
  \end{tabular}
     \caption{Examples of waveform encrypted by Flipping and ROM}
     \label{bit-orth-wav-spe}
\end{figure}

\begin{figure}[t]
\centering
  \begin{tabular}{ccc}
  \begin{minipage}[b]{0.28\columnwidth}
    \centering
    \includegraphics[keepaspectratio,width=\columnwidth]{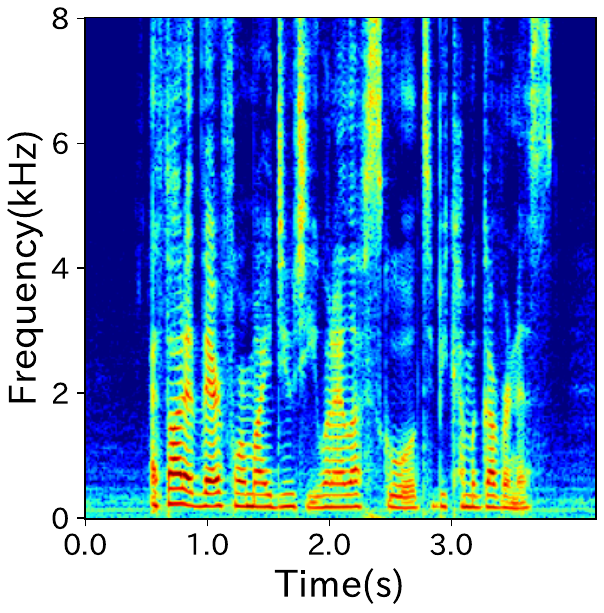}
    \subcaption{Original spectrogram}\vspace{1.7em}
    \label{fig:original-spe-pix}
  \end{minipage} &
      \begin{minipage}[b]{0.28\columnwidth}
    \centering
    \includegraphics[keepaspectratio, width=\columnwidth]{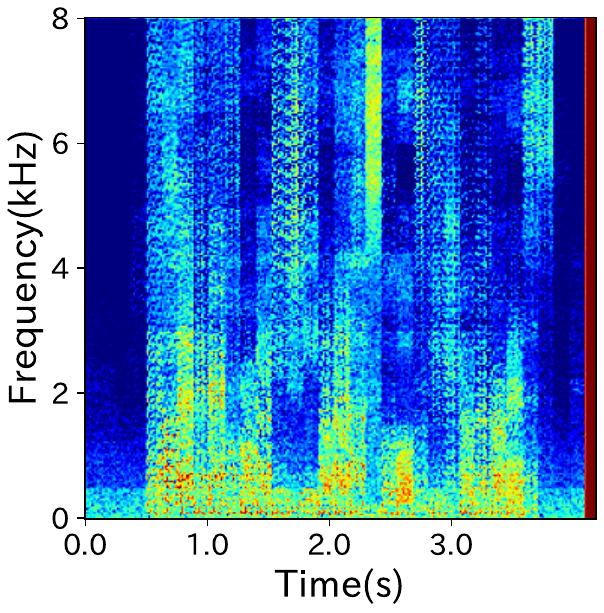}
    \subcaption{Encrypted spectrogram\\ ($M=8$) }
    \label{fig:pix8-spe}
  \end{minipage}&
  \begin{minipage}[b]{0.28\columnwidth}
    \centering
    \includegraphics[keepaspectratio, width=\columnwidth]{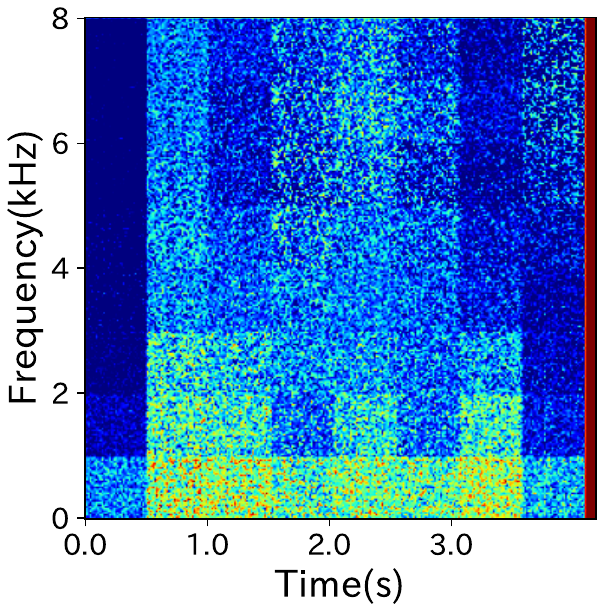}
    \subcaption{Encrypted spectrogram\\ ($M=32$) }
    \label{fig:pix32-spe}
  \end{minipage}
  
  \end{tabular}
     \caption{Examples of spectrogram encrypted by Shuffling}
     \label{figs:pix-spes}
\end{figure}

\begin{figure}[t]
\centering
  \begin{tabular}{cc|cc}
  \begin{minipage}[b]{0.2\columnwidth}
    \centering
    \includegraphics[keepaspectratio, width=\columnwidth]{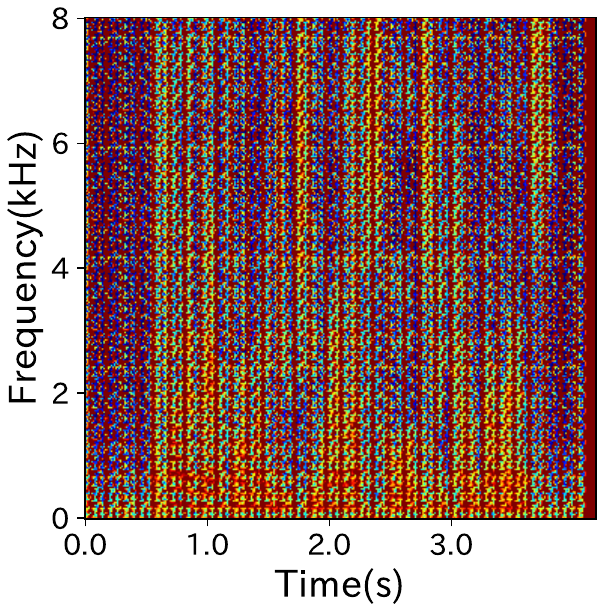}
    \subcaption{Encrypted spectrogram\\ ($M=8$) }
    \label{fig:bit8-spe}
  \end{minipage}&
  \begin{minipage}[b]{0.2\columnwidth}
    \centering
    \includegraphics[keepaspectratio, width=\columnwidth]{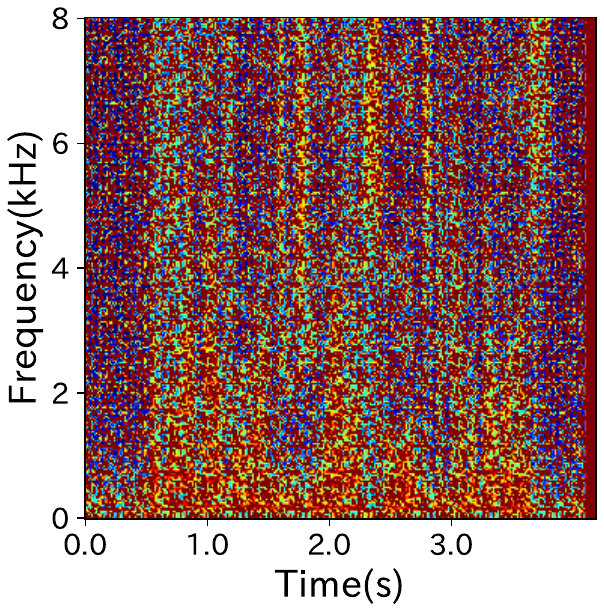}
    \subcaption{Encrypted spectrogram\\ ($M=32$) }
    \label{fig:bit32-spe}
  \end{minipage}&
  \begin{minipage}[b]{0.2\columnwidth}
    \centering
    \includegraphics[keepaspectratio, width=\columnwidth]{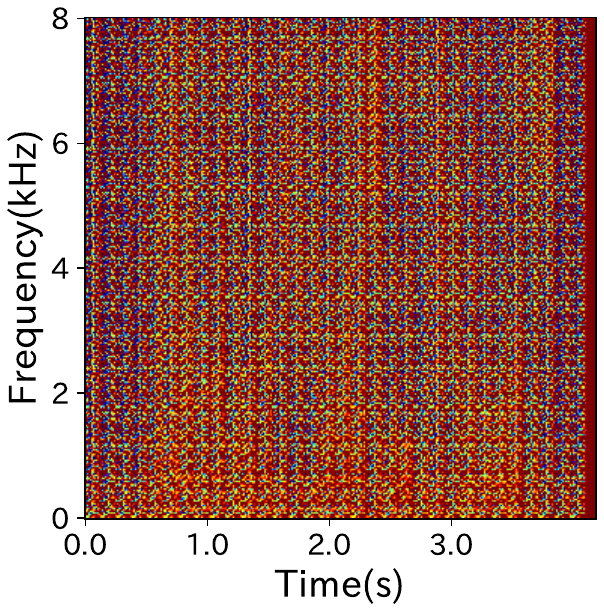}
    \subcaption{Encrypted spectrogram\\ ($M=8$) }
    \label{fig:orth8-spe}
  \end{minipage}&
  \begin{minipage}[b]{0.2\columnwidth}
    \centering
    \includegraphics[keepaspectratio, width=\columnwidth]{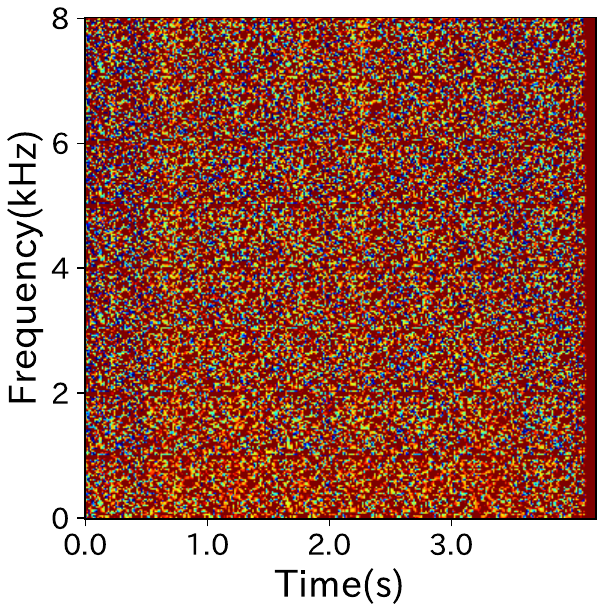}
    \subcaption{Encrypted spectrogram\\ ($M=32$) }
    \label{fig:orth32-spe}
  \end{minipage}\\
  \multicolumn{2}{c}{\footnotesize \textbf{Flipping}} & \multicolumn{2}{c}{\footnotesize \textbf{ROM}}\\
  \end{tabular}
     \caption{Examples of spectrogram encrypted by Flipping and ROM}
     \label{figs:bit-orth-spes}
\end{figure}

\subsection{Comparison of encrypted speech data}
In this section, we investigated the characteristics of each proposed encryption method in some examples.
The original speech waveform is shown in Fig.~\ref{pix-ori-wav}, Figs.~\ref{pix10-wav} and \ref{pix128-wav} are waveforms encrypted by Shuffling from the original waveform with different key sizes $(M=10, 128)$, and Figs.~\ref{pix-ori-wav-spe}, \ref{pix10-wav-spe}, and \ref{pix128-wav-spe} are their spectrograms, respectively.
Figure~\ref{bit-orth-wav-spe} shows the waveforms encrypted by Flipping and ROM and their spectrograms. The key sizes are the same as the Flipping case.
From these figures, it can be confirmed that the speech waveforms encrypted by the proposed encryption methods are significantly different from the original speech waveform.
The spectrograms of each speech waveform show that the original speech waveform is significantly different due to the encryption using the proposed methods, and the characteristics of the original speech waveform are also significantly different.
It was also found that the larger the value of the block size~$M$ used for the proposed encryption methods, the larger the change in the waveforms.
With a larger value of $M$, the speech content can be concealed more effectively. Additionally, from the perspective of key space, a larger value of $M$ makes it more difficult to estimate the secret key.

The original speech waveform is shown in Fig.~\ref{fig:original-spe-pix}, and Figs.~\ref{fig:pix8-spe} and \ref{fig:pix32-spe} are spectrograms encrypted from the original spectrogram by Shuffling with different key sizes $(M=8, 32)$.
By comparing Figs.~\ref{fig:original-spe-pix} and~\ref{fig:pix8-spe} it can be seen that the harmonic structures of each spectrogram are distorted.
Comparing Figs.~\ref{fig:pix8-spe} and \ref{fig:pix32-spe}, we can see that the larger the value of $M$, the greater the range of movement for the positions of values in each block of the spectrograms. 
The spectrograms encrypted by Flipping and ROM are shown in Fig.~\ref{figs:bit-orth-spes}.
By comparing Figs.~\ref{fig:original-spe-pix} and~\ref{fig:bit8-spe}, and Figs.~\ref{fig:original-spe-pix} and~\ref{fig:orth8-spe}, we can see that the magnitude of each value in the encrypted spectrogram changes randomly.
Comparing Figs.~\ref{fig:bit8-spe} and \ref{fig:bit32-spe}, and Figs.~\ref{fig:orth8-spe} and \ref{fig:orth32-spe}, we can see that the spectrogram values in the block change regardless of the block size~$M$.
From these figures, it can be visually confirmed that the encryption of the proposed methods succeeded in anonymizing the speech with encryption since the harmonic structure of the original spectrogram was hidden and, in particular, ROM obscured the speech segment and other information.

\begin{figure}[t]
\centering
  \begin{tabular}{ccc}
  \begin{minipage}[b]{0.28\columnwidth}
    \centering
    \includegraphics[keepaspectratio,width=\columnwidth]{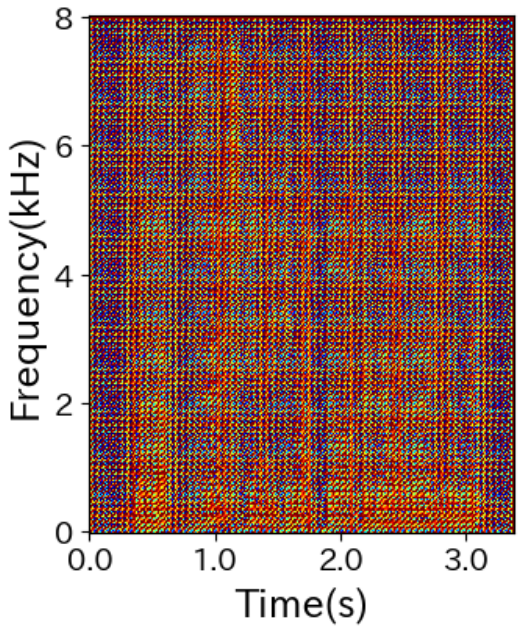}
    \subcaption{Encrypted spectrogram\\ $(M=3)$}
    \label{fig:orth3_wc}
  \end{minipage} &
      \begin{minipage}[b]{0.28\columnwidth}
    \centering
    \includegraphics[keepaspectratio, width=\columnwidth]{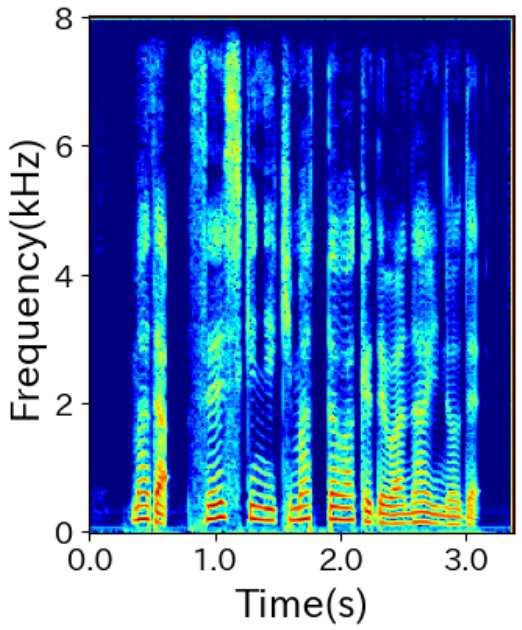}
    \subcaption{Spectrogram decrypted with correct key}
    \label{fig:orth3_correct}
  \end{minipage}&
  \begin{minipage}[b]{0.28\columnwidth}
    \centering
    \includegraphics[keepaspectratio, width=\columnwidth]{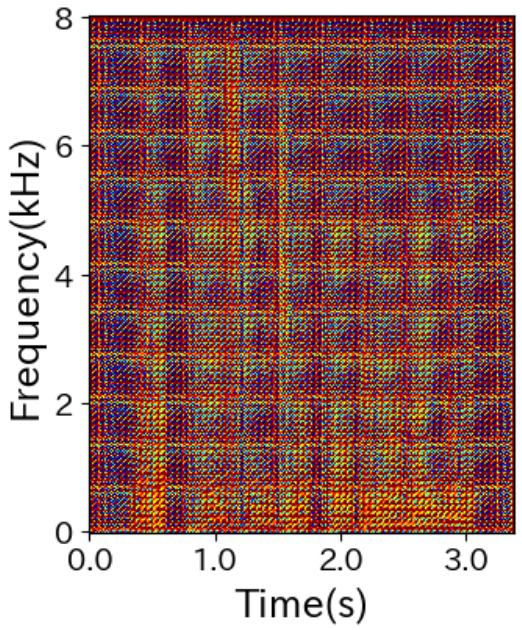}
    \subcaption{Spectrogram decrypted with incorrect key}
    \label{fig:orth3_wrong}
  \end{minipage}
  
  \end{tabular}
  \caption{Examples of decryption for spectrograms encrypted by ROM under $M=3$.}
  \label{figs:orth_wc}
\end{figure}

\begin{figure}[t]
\centering
  \begin{minipage}[b]{0.3\linewidth}
    \centering
    \includegraphics[keepaspectratio,width=\linewidth]{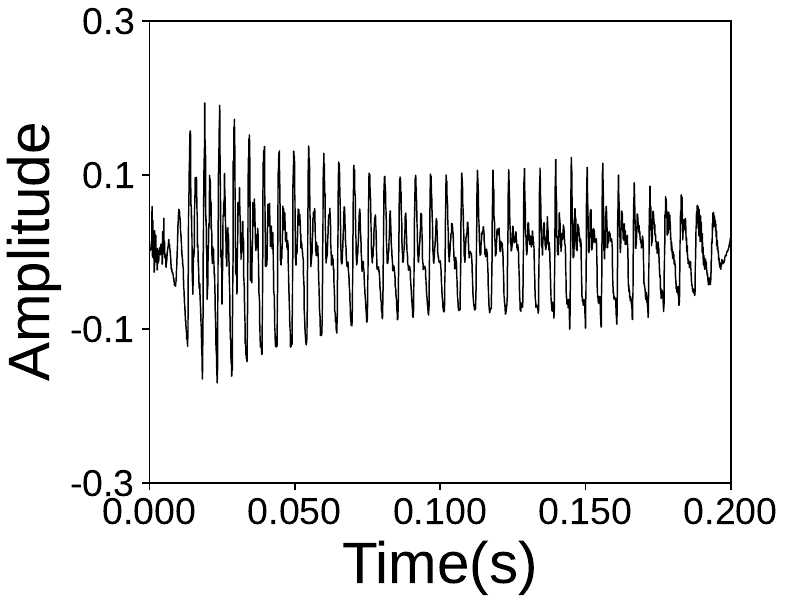}
    \includegraphics[keepaspectratio,width=\linewidth]{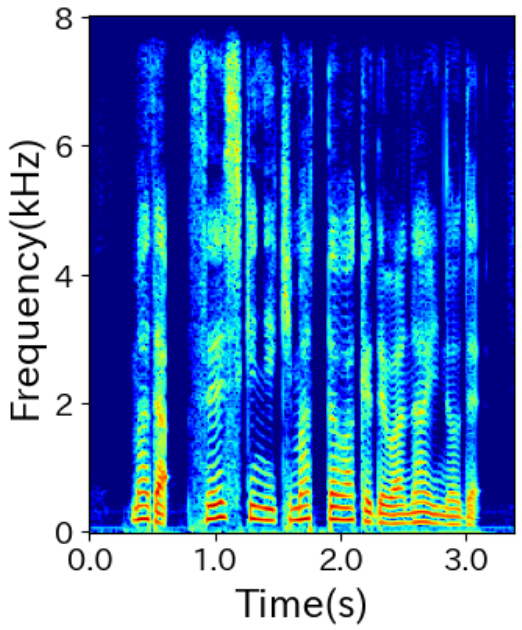}
    \subcaption{Original}\vspace{1em}
    \label{fig:ori}
  \end{minipage}
  \hspace{0.02\columnwidth}
  \begin{minipage}[b]{0.3\linewidth}
    \centering
    \includegraphics[keepaspectratio, width=\linewidth]{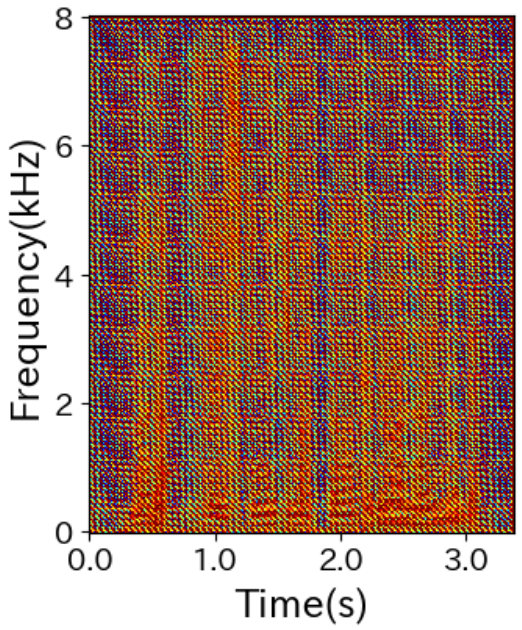}
    \subcaption{Encrypted spectrogram\\ $(M=3)$}
    \label{fig:orth3}
  \end{minipage}
  \hspace{0.02\columnwidth}
  \begin{minipage}[b]{0.3\linewidth}
    \centering
    \includegraphics[keepaspectratio,width=\linewidth]{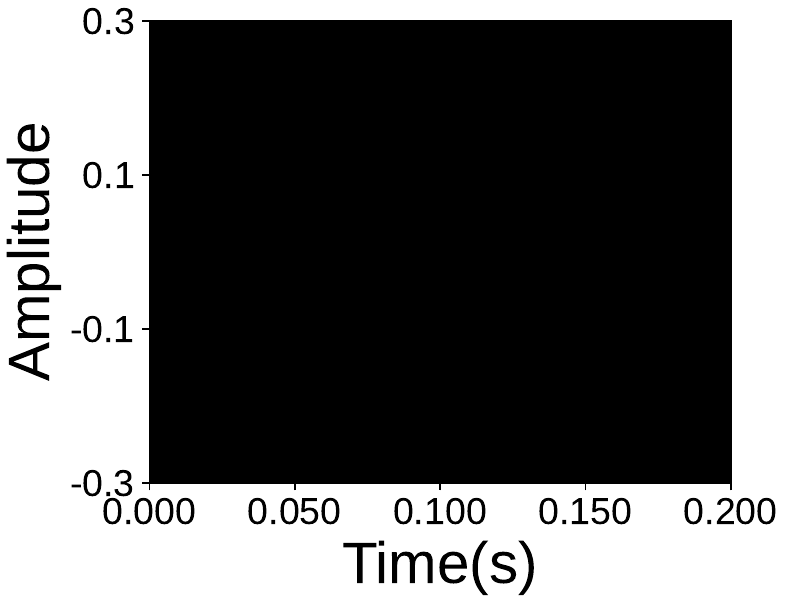}
    \includegraphics[keepaspectratio, width=\linewidth]{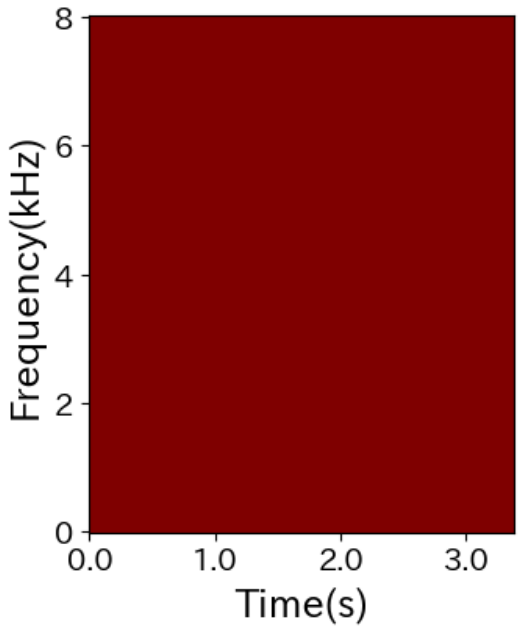}
    \subcaption{Phase reconstruction $(M=3)$}
    \label{fig:orth3re}
  \end{minipage}
  \caption{Example of phase reconstruction applied to encrypted spectrogram by ROM under $M=3$.}
  \label{figs:orth} 
\end{figure}

\begin{figure}[t]
    \centering
    \includegraphics[keepaspectratio,width=8cm]{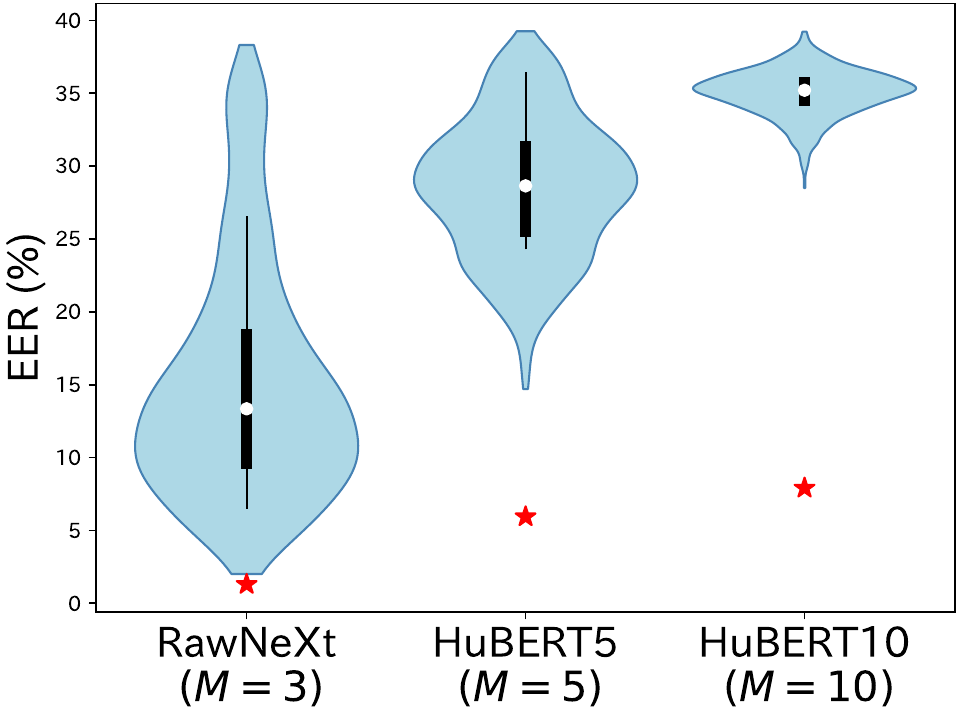}
    \caption{Distribution of EER (\%) for speech waveforms encrypted using $1000$ ROM secret keys and input to encrypted ASV model (asterisk: EER (\%) for unencrypted ASV model)} 
    \label{fig:asv-eer}
\end{figure}

\begin{figure}[t]
    \centering
    \includegraphics[keepaspectratio,width=8cm]{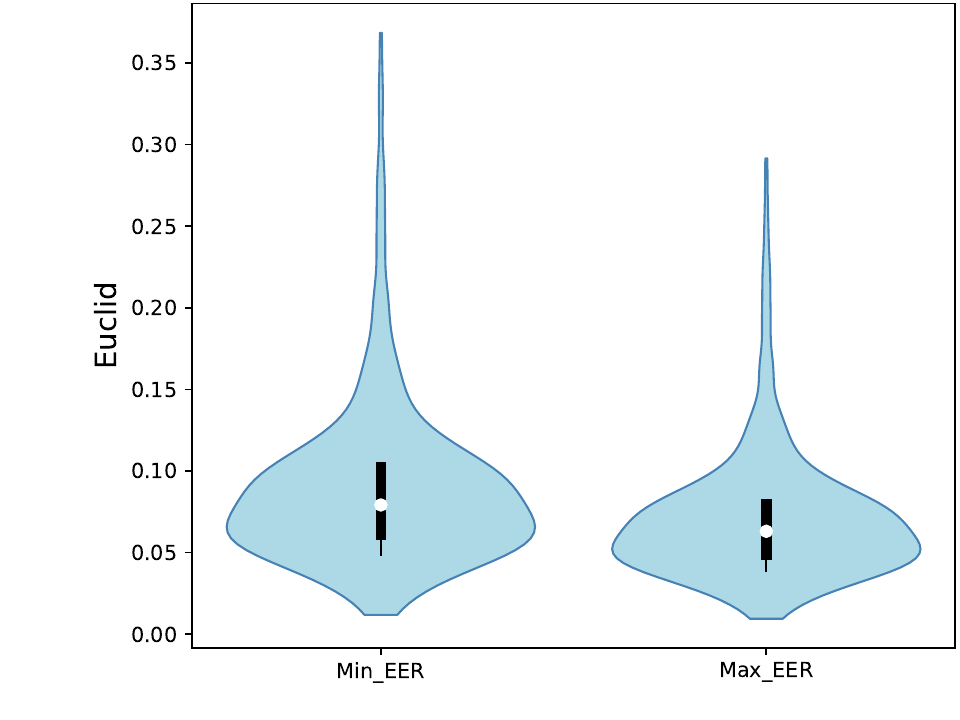}
    \caption{Distribution of Euclidean distance of speech waveforms encrypted with wrong ROM secret keys and speech waveforms encrypted with correct keys} 
    \label{fig:asv-eer-key-mse}
\end{figure}

\begin{figure}[t]
    \centering
    \includegraphics[keepaspectratio,width=7.5cm]{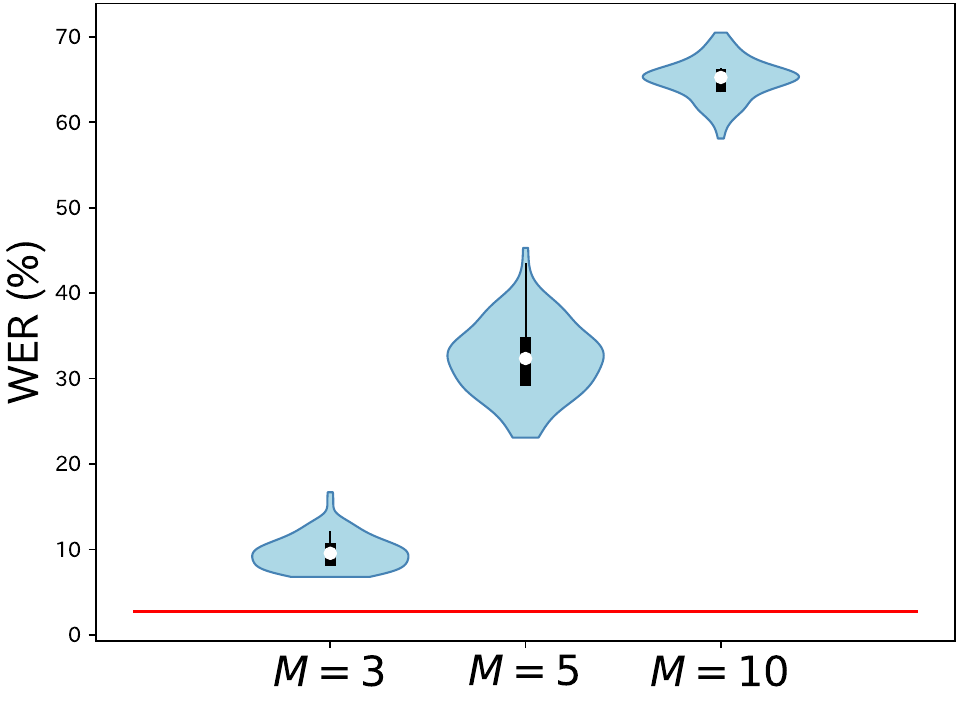}
    \caption{Distribution of WER$(\%)$ in case of speech waveforms encrypted using $100$ ROM secret keys and input to ASR model~\cite{ESPnetASR} (line: WER $(\%)$ for unencrypted speech waveforms)} 
    \label{fig:asr-wer}
\end{figure}

\subsection{Robustness evaluation experiments}
\subsubsection{Experimental conditions}
Four experiments were conducted to evaluate robustness against attacks with the proposed methods.
In the first experiment, we investigated the changes in the spectrogram encrypted by ROM when decrypted with the correct key and an incorrect key.
In the second experiment, the spectrogram encrypted by ROM was reconstructed into a speech waveform using Phase Gradient Heap Integration~(PGHI)~\cite{pghi2017}, one of the state-of-the-art phase reconstruction algorithms, and 
the reconstructed speech waveform was compared with the original one.
In the third experiment, we prepared an ASV model encrypted with the correct secret key and input speech encrypted with $1000$ incorrect keys to the model.
The two ASV models used in this experiment were HuBERT~\cite{hubert} and RawNeXt~\cite{rawnext}.
As in the experiment in Section~\ref{seq:experiment}, we prepared two kinds of HuBERT model that included the first convolutional layer with kernel sizes and stride sizes of five and ten, respectively.
The first convolutional layers of HuBERT5 and HuBERT10 were encrypted by ROM.
The RawNeXt model took speech waveforms as input features and was trained using the VoxCeleb2 corpus~\cite{voxceleb2}.
In this experiment, encryption with ROM was applied to the pre-trained models distributed in~\cite{rawnext}.
The RawNeXt model was originally set to a stride size and kernel size of three.
For evaluation, we used the VoxCeleb1 corpus's test set~\cite{voxceleb}, and EER was used as the evaluation measure.
In the fourth experiment, the same experiment as the third experiment was applied to the ASR task.
In the experiments, speech waveforms encrypted with $100$ ROM secret keys were input to a trained ASR model proposed in the~\cite{ESPnetASR}.
Under $M=3,5,10$, we used the test clean subset of LibriSpeech as the evaluation data.

\subsubsection{Experimental results}
Figure~\ref{figs:orth_wc} shows the result of the first experiment of the robustness evaluation.
Figure~\ref{fig:orth3_wc} is the spectrogram encrypted by using ROM from the original spectrogram in Fig.~\ref{pix-ori-wav-spe} under block size $M = 3$. 
Figure~\ref{fig:orth3_correct} shows the spectrogram in Fig.~\ref{fig:orth3_wc} decrypted using the correct key, while Fig.~\ref{fig:orth3_wrong} shows the spectrogram in Fig.~\ref{fig:orth3_wc} decrypted using the incorrect key.
Figures~\ref{pix-ori-wav-spe} and \ref{fig:orth3_correct} show that the spectrogram decrypted with the correct key was exactly the same as the original spectrogram. 
On the other hand, Fig.~\ref{fig:orth3_wrong} shows that the original spectrogram information was not decrypted when an incorrect key was used as a decrypted key.
Figures~\ref{pix-ori-wav-spe} and \ref{fig:orth3_correct} show that the spectrograms were exactly the same as the original spectrograms when decrypted with the correct keys.
On the other hand, Figs.~\ref{pix-ori-wav-spe} and \ref{fig:orth3_wrong} show that the original spectrogram information was not decrypted when decrypted using an incorrect key.

The results of phase reconstruction using PGHI on the spectrogram encrypted using ROM are shown in Fig.~\ref{figs:orth} as the second experiment of the robustness evaluation.
Figure~\ref{fig:ori} is the original spectrogram, Fig.~\ref{fig:orth3} is the spectrogram obtained by encrypting the spectrogram in Fig.~\ref{fig:ori} with ROM under block size $M=3$, and Fig.~\ref{fig:orth3re} is the spectrogram of the speech waveform obtained by applying PGHI to the spectrogram in Fig.~\ref{fig:orth3}.
The figures shown in the upper rows of Figs.~\ref{fig:ori} and \ref{fig:orth3re} are the original speech waveform and the speech waveform obtained by PGHI, respectively.
Comparing Fig.~\ref{fig:ori} with \ref{fig:orth3re}, it can be seen that the structure of the original spectrogram was not reconstructed in the spectrogram after PGHI.
In addition, comparing the waveforms shown in the upper rows of Fig.~\ref{fig:ori} and Fig.~\ref{fig:orth3re}, it can be seen that the original waveforms were not reconstructed at all.
Therefore, it is confirmed that a spectrogram encrypted using the proposed methods can hardly reconstruct the original speech when the correct key is not known.

For the third experiment in the robustness evaluation, we analyzed how the proposed method behaves with incorrect keys in the ASV task.
A violin plot of the distribution of EER when $1000$ ROM secret keys were used as incorrect keys is shown in Fig.~\ref{fig:asv-eer}.
The EER for each model, when the speech was not encrypted, is represented by stars in Fig.~\ref{fig:asv-eer}.
The variance of EER for RawNeXt was $66.2$, that for HuBERT5 was $24.1$, and that for HuBERT10 was $2.51$. 
Figure~\ref{fig:asv-eer} shows that the smaller the block size used for encryption, the wider the EER distribution, and it also shown that the larger the block size, the larger the difference from the EER with the correct key.
The reason why the distribution of RawNeXt's EER is biased toward low positions can be considered to be the small block size and the high generalization performance of the model.
Furthermore, to investigate the characteristics of the generated keys, we measured the Euclidean distance between the speech encrypted with the correct key and the speech encrypted with incorrect keys. We presented the distributions for key groups with low EER and high EER in Figure~\ref{fig:asv-eer-key-mse}. From this figure, we observed that keys with a low EER are not necessarily closer to the audio encrypted with the correct key. The characteristics of the secret keys should be investigated in the future.

For the fourth experiment in the robustness evaluation, we analyzed how ROM behaves with incorrect keys in the ASR task.
Figure~\ref{fig:asr-wer} shows the distribution of WER when speech waveforms encrypted with $100$ ROM secret keys were input to the trained speech recognition model proposed in \cite{ESPnetASR}.
The red line in Fig.~\ref{fig:asr-wer} is the WER when plain speech was input to the plain model.
As well as with the results of the third experiment, as the value of $M$ increased, the more the average WER rose, and the more the distribution became narrower. 
This implies that a larger value of $M$ results in higher confidentiality for speech content.
The ASV and ASR results confirm that the proposed methods provide stable privacy-preserving performance when the block size is large.

\section{Conclusion}\label{seq:conclude}
In this paper, we described privacy-preserving methods using secret keys based on Shuffling, Flipping, and ROM.
The proposed methods can perform convolutional computation to cancel the effect of the orthogonal matrix secret key so that encrypted queries can be input to encrypted CNN models without decryption.
In addition, when the user knows the correct key, there is no performance degradation at all.
Experiments confirmed that, when using the proposed methods, users who do not know the secret key used to encrypt the model cannot use the model with high performance, and the larger the block size used for encryption, the more stable the privacy-preserving performance is.
It was also confirmed that a third party who does not know the secret key cannot estimate or reconstruct the original speech data from the speech data encrypted using the proposed methods.
For future work, we will develop an encryption method for speech data that is stable and robust against attacks, even when the block size is small, and we will investigate the effect of some noise reduction methods to confuse the inner product calculations.
In addition, it is an important topic to expand our approach not only to CNN-based models but also to deep-learning models.
As a challenging and crucial task, we also plan to explore research focusing on effectively concealing a part of critical components.

\section*{Acknowledgment}
This work was supported in part by JSPS KAKENHI (Grant Number JP21H01327), JST CREST (Grant Number JPMJCR20D3) 
, and ROIS DS-JOINT (047RP2023) to S. Shiota.

\bibliographystyle{IEEEtran} 
\bibliography{ref} 

\end{document}